\documentclass{article}
\pdfoutput=1
\usepackage{lscape}
\usepackage{enumerate}
\usepackage{amsfonts}
\usepackage{amsmath}
\usepackage{textcomp}
\usepackage{amssymb}
\usepackage{color}
\usepackage{graphicx}
\usepackage{pdflscape}
\usepackage{afterpage}
\usepackage[matrix,arrow,curve]{xy}
\usepackage{array}
\usepackage{indentfirst}
\usepackage{subcaption}

\def\be{\begin{equation}}
\def\ee{\end{equation}}

\makeatletter
\renewcommand*{\@cite@ofmt}{\bfseries\hbox}
\makeatother

\def\green{\color{green}}
\def\blue{\color{blue}}
\def\red{\color{red}}
\def\black{\color{black}}

\hoffset= - 1.2in         

\begin{document}

\title{\vspace{0.1cm}{\Large {\bf  Entangling gates from cabling of knots}\vspace{.2cm}}
\author{\bf Sergey Mironov$^{a,c,d}$\thanks{e-mail: sa.mironov\_1@physics.msu.ru},
Andrey Morozov$^{b,c,e}$\thanks{e-mail: morozov.andrey.a@iitp.ru}}
}
\date{ }

\maketitle

\vspace{-5.5cm}

\begin{center}
\hfill ITEP/TH-41/24\\
\hfill IITP/TH-35/24\\
\hfill MIPT/TH-25/24\\
\end{center}

\vspace{3.6cm}

\begin{center}
$^a$ {\small {\it INR RAS, Moscow 117312, Russia}}\\
$^b$ {\small {\it IITP RAS, Moscow 127994, Russia}}\\
$^c$ {\small {\it NRC ``Kurchatov Institute'', Moscow 123182, Russia}}\\
$^d$ {\small {\it ITMP, MSU, Moscow, 119991, Russia}}\\
$^e$ {\small {\it MIPT, Dolgoprudny, 141701, Russia}}\\
\end{center}

\vspace{1cm}

\begin{abstract}
While there is a general consensus about the structure of one qubit operations in topological quantum computer, two qubits are as usual a more difficult and complex story of different attempts with varying approaches, problems and effectiveness. In this paper we discuss how to construct an efficient realization of a two qubit gate in topological quantum computer, by using principle of cabling from the knot theory. This allows oneto construct a braiding of cables dependent on the parameters of the theory where there is a low probability of moving out of computational space (high fidelity of operation) with a non-trivial entangling two-qubit operation. We also present some examples of these operations for different parameters of the theory.
\end{abstract}

\vspace{.5cm}



\section{Introduction}
Quantum computer is an effective tool to solve some problems that on usual computer would require exponentially large resources. While there are many particular technical realizations of quantum computers and devices, they all are based on conventional quantum systems/effects and are plagued with noise and errors. One of the ways to completely avoid random mistakes in the algorithm is to consider topological quantum computer \cite{Kit}. States and operations in topological quantum computer (TQC) are realized by quasi-particles and their evolution in some topological field theory. The topological nature of the theory protects the algorithm against random errors. Today TQC still lacks the experimental realization of non-abelian anyons, but extensive research is performed in this direction \cite{ReviewTQC,exp1,exp2,exp3,exp4,exp5,expend}.

The natural choice for the model of non-Abelian anyons and TQC is 3d Chern-Simons theory with gauge group $SU(N)$:
\begin{equation}
S_{CS}=\frac{k}{4\pi}\int\limits_{S^3}\text{Tr}\left( A\wedge dA+\frac{2}{3}A\wedge A\wedge A\right),
\label{e:SCS}
\end{equation}
where $k$ is an integer coupling constant.

States of anyons and anti-anyons correspond to particular states of qubits, while intertwining of their world-lines in a three-dimensional space can be seen as operations as they change the states of the particles.

Observables in quantum Chern-Simons theory possess a more complex symmetry than a $SU(N)$ we put in the action. Namely they display properties, characteristic to the quantum group (quantum algebra) $U_q(SU(N))$, \cite{Klimyk}. It is a deformation of universal enveloping algebra of the $SU(N)$ and is widely studied in relation to the different models in theoretical and mathematical physics. In the context of this paper it is important that the crossings of the world lines correspond to the quantum $\mathcal{R}$-matrices of $U_q(SU(N))$ \cite{KLJ}.

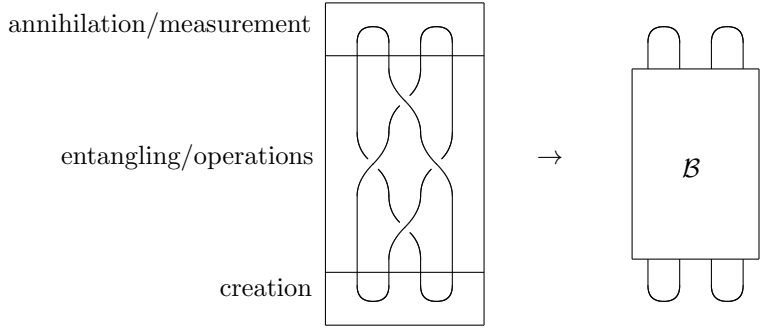
\begin{figure}[h!]
\begin{picture}(150,140)(-230,-65)
\qbezier(-12,0)(-18,-6)(-18,-12)
\qbezier(-10,-2)(-6,-6)(-6,-12)
\qbezier(10,-2)(6,-6)(6,-12)
\qbezier(12,0)(18,-6)(18,-12)
\put(-18,-12){\line(0,-1){34}}
\qbezier(-6,-12)(-6,-18)(-2,-22)
\qbezier(6,-12)(6,-18)(0,-24)
\put(18,-12){\line(0,-1){34}}
\qbezier(0,-24)(-6,-30)(-6,-36)
\qbezier(2,-26)(6,-30)(6,-36)
\put(-6,-36){\line(0,-1){10}}
\put(6,-36){\line(0,-1){10}}
\qbezier(-18,-46)(-18,-52)(-12,-52)
\qbezier(-6,-46)(-6,-52)(-12,-52)
\qbezier(18,-46)(18,-52)(12,-52)
\qbezier(6,-46)(6,-52)(12,-52)
\qbezier(-14,2)(-18,6)(-18,12)
\qbezier(-12,0)(-6,6)(-6,12)
\qbezier(12,0)(6,6)(6,12)
\qbezier(14,2)(18,6)(18,12)
\put(-18,12){\line(0,1){34}}
\qbezier(-6,12)(-6,18)(-2,22)
\qbezier(6,12)(6,18)(0,24)
\put(18,12){\line(0,1){34}}
\qbezier(0,24)(-6,30)(-6,36)
\qbezier(2,26)(6,30)(6,36)
\put(-6,36){\line(0,1){10}}
\put(6,36){\line(0,1){10}}
\qbezier(-18,46)(-18,52)(-12,52)
\qbezier(-6,46)(-6,52)(-12,52)
\qbezier(18,46)(18,52)(12,52)
\qbezier(6,46)(6,52)(12,52)
\put(-30,41){\line(1,0){60}}
\put(-30,61){\line(0,-1){122}}
\put(30,-41){\line(-1,0){60}}
\put(30,-61){\line(0,1){122}}
\put(-30,61){\line(1,0){60}}
\put(-30,-61){\line(1,0){60}}
\put(-150,50){\hbox{annihilation/measurement}}
\put(-130,0){\hbox{entangling/operations}}
\put(-70,-50){\hbox{creation}}
\put(50,0){\hbox{$\rightarrow$}}
\put(110,0){
\put(-18,-36){\line(0,-1){10}}
\put(18,-36){\line(0,-1){10}}
\put(-6,-36){\line(0,-1){10}}
\put(6,-36){\line(0,-1){10}}
\qbezier(-18,-46)(-18,-52)(-12,-52)
\qbezier(-6,-46)(-6,-52)(-12,-52)
\qbezier(18,-46)(18,-52)(12,-52)
\qbezier(6,-46)(6,-52)(12,-52)
\put(-18,36){\line(0,1){10}}
\put(18,36){\line(0,1){10}}
\put(-6,36){\line(0,1){10}}
\put(6,36){\line(0,1){10}}
\qbezier(-18,46)(-18,52)(-12,52)
\qbezier(-6,46)(-6,52)(-12,52)
\qbezier(18,46)(18,52)(12,52)
\qbezier(6,46)(6,52)(12,52)
\put(-24,-36){\line(1,0){48}}
\put(-24,-36){\line(0,1){72}}
\put(24,36){\line(-1,0){48}}
\put(24,36){\line(0,-1){72}}
\put(-5,-5){\hbox{$\mathcal{B}$}}
}
\end{picture}
\caption{Description of one-qubit operations using anyons. Two pairs of anyons are created then they are entangled and then annihilated. Trajectories of anyons form a knot or a link. The braid/entangling of two pairs of anyons corresponds to an operator $\mathcal{B}$ which is a two by two matrix.\label{f:anyons}}
\end{figure}

One of the possibilities to define the computational space of the qubit, as was extensively discussed in \cite{TopoBook,TowTQC,QuantKnots,LargeK,Measure,2qArb}, is to consider 4-plat representation of knot or link where state of the qubit corresponds to the space of irreducible representations in two anyon-anti-anyon pairs (see Fig.\ref{f:anyons} for an example). We discuss the explicit construction in the next section.

Previously it was proven that quantum $\mathcal{R}$-matrices form a universal set of one-qubit gates at least for large enough values of $k$ \cite{QuantKnots,LargeK}. As usual multi-qubit and two-qubit operations prove to be difficult to implement efficiently. In the case of the topological quantum computer the main difficulty is that there is a possibility that these operations can move the system out of the computational space \cite{TopoBook}. Realization suggested in this paper also has this problem, however, we claim that with accurate choice of operation this probability can be made sufficiently small. But to achieve this the operation will depend on the particular realization of TQC, i.e. on the values of $k$ and $N$.


\section{One-qubit operations \label{s:1qb}}

The initial state, that is depicted in a lower block of Fig.\ref{f:anyons} corresponds to the creation of two anyon pairs. Upon creation two anyons in each pair are in conjugate representations, and pair as a whole is projected onto trivial representation. After intertwining, which corresponds to an action of $\mathcal{R}$-matrices, the representation of a pair can change. Still, braid as a whole remains in trivial representation, since $\mathcal{R}$-matrices do not change the irreducible representations travelling along any number of strands. We assume for simplicity, that each strand is in fundamental/anti-fundamental representation, however a particular experimental realization may be built on anyons in different representations which would depend on the types of anyons we would be able to create. As it was previously suggested, \cite{TopoBook,TowTQC,QuantKnots,LargeK,Measure,2qArb} the 4-plat knot (also known as two-bridge knots), i.e. knot with two initial pairs of anyons, corresponds to one qubit computing. Indeed, the projection onto trivial representation of the product of two fundamental and two anti-fundamental representations is two-dimensional. To get the trivial representation in the whole braid, each of the two pairs of strands should be either in trivial or adjoint representation. It can be seen, for example, as follows:
\begin{equation}
\label{1q}
\begin{array}{l}
|V_{\emptyset}\rangle:\ \ \ ([1]\otimes\bar{[1]})\otimes([1]\otimes\bar{[1]})\rightarrow\emptyset\otimes\emptyset\rightarrow\emptyset,
\\
|V_1\rangle:\ \ \ ([1]\otimes\bar{[1]})\otimes([1]\otimes\bar{[1]})\rightarrow\text{adj}\otimes\text{adj}\rightarrow\emptyset.
\end{array}
\end{equation}

Vectors $|V_{\emptyset}\rangle$ and $|V_1\rangle$ form a Hilbert space of one qubit. Let us stress, that computational space of a qubit is the space of different irreducible representations rather than an internal vector space of a particular irreducible representation. The closest analogy are $SU(3)$ and isospin multiplets of particles. Anyons in different representations correspond to particles in different hypermultiplets, in other words, to different particles.

We define four two-dimensional operators, acting on a qubit \cite{QuantKnots,Tabul,DoubleFat,NPZ,MMS}:
 \begin{equation}
\begin{array}{lcrclcr}
S&=&
\frac{1}{\sqrt{(q+q^{-1})(A-A^{-1})}}\left(\begin{array}{cc}\sqrt{\frac{A}{q}-\frac{q}{A}} & \sqrt{Aq-\frac{1}{Aq}} \\ \sqrt{Aq-\frac{1}{Aq}} & -\sqrt{\frac{A}{q}-\frac{q}{A}} \end{array}\right);
& &
T&=&\left(\begin{array}{cc}
\frac{q}{A} \\ & -\frac{1}{qA}
\end{array}\right);
\\ \\
\bar{S}&=&
\left(\begin{array}{cc}\frac{q-q^{-1}}{A-A^{-1}} & \frac{\sqrt{(Aq-\frac{1}{Aq})(\frac{A}{q}-\frac{q}{A})}}{A-A^{-1}} \\ \frac{\sqrt{(Aq-\frac{1}{Aq})(\frac{A}{q}-\frac{q}{A})}}{A-A^{-1}} & -\frac{q-q^{-1}}{A-A^{-1}} \end{array}\right);
 & &
\bar{T}&=&\left(\begin{array}{cc}
1 \\ & -A
\end{array}\right),
\end{array}
\label{eq:ST}
\end{equation}
where $q=\text{exp}(\frac{2\pi i}{k+N})$ and $A=q^N$ with $k$ and $N$ being parameters of Chern-Simons theory (\ref{e:SCS}).
In the $SU(N)$ case anyons and anti-anyons are different particles and therefore operators for the crossings of the trajectories depend on the exact type of particles which are crossing. If they are of the same type then operator $T$ should be chosen, if they are different then the operator is $\bar{T}$. $T$ and $\bar{T}$ are diagonalized $\mathcal{R}$-matrices written for the particular representations. Also since we write operators which act on all four anyons, we cannot diagonalize all of the at the same time. Matrices $T$ and $\bar{T}$ are diagonalized operators, while operators $S$ and $\bar{S}$ describe the change of bases to make different crossing operators diagonal. They also depend on the placement of particles of different types therefore there are two of them. See \cite{QuantKnots,LargeK} for more details. As was discussed in \cite{QuantKnots,LargeK} these operators form a universal set of one-qubit gates (universality holds for qudits upon generalization to higher representations).

Thereby, after two pairs of anyon and anti-anyon are created (the initial state of the quantum computer), operations on a computational space are performed with operators (\ref{eq:ST}) in accordance with particular braid: one should take a product of operators corresponding to the different crossings in an order defined by a braid. Last part is the annihilation of anyon-anti-anyon pairs\footnote{Note that due to intersections, the annihilated particles in each pair might differ from initial ones.} which corresponds to the projection onto trivial representation of each pair of strands. The process of annihilation has non-trivial probability, which is the result of a quantum program. From the representations point of view, braid (algorithm) corresponds to a two-by-two matrix\footnote{Let us remind that $X$ and $Y$ takes values in space of different irreducible representations.} $\mathcal{B}_{XY}$, which is a product of matrices from (\ref{eq:ST}). The full program with initial creation and measurement corresponds to a matrix element $\mathcal{B}_{11}$, when $X=Y=\emptyset$, it describes the probability amplitude of two pairs of anyons being created, then entangled and then annihilated.

\section{Two-qubits and the entangling operation}

As we saw in section \ref{s:1qb}, one plat that contains four strands represents a qubit, which computational space is two dimensional. Indeed, each of two pairs of strands is in superposition of {\it trivial} and {\it adjoint} representations, which gives four possible combinations. But after projection of the whole plat to the trivial representations (which is implied by a construction) only two combinations survive: $\text{adj}\otimes\text{adj}$ and $\emptyset\otimes\emptyset$. If we take two plats together, it can seem that there should be four dimensional space. And it would be the case as long as two plats are not intertwined with each other, see Fig.\ref{2free}
\begin{figure}[h!]
\begin{subfigure}{.3\textwidth}
\begin{picture}(150,190)(-30,-75)
\put(0,-50){
\put(0,0){\line(1,0){25}}
\put(0,0){\line(0,1){35}}
\put(25,35){\line(-1,0){25}}
\put(25,35){\line(0,-1){35}}
\put(5,13){\hbox{$\mathcal{B}^{(1)}_{1}$}}
\qbezier(3,0)(3,-8)(6,-8)
\qbezier(9,0)(9,-8)(6,-8)
\qbezier(16,0)(16,-8)(19,-8)
\qbezier(22,0)(22,-8)(19,-8)
}
\put(0,76){
\put(0,0){\line(1,0){25}}
\put(0,0){\line(0,1){35}}
\put(25,35){\line(-1,0){25}}
\put(25,35){\line(0,-1){35}}
\put(5,13){\hbox{$\mathcal{B}^{(1)}_{n}$}}
\qbezier(3,35)(3,43)(6,43)
\qbezier(9,35)(9,43)(6,43)
\qbezier(16,35)(16,43)(19,43)
\qbezier(22,35)(22,43)(19,43)
}
\put(36,-50){
\put(0,0){\line(1,0){25}}
\put(0,0){\line(0,1){35}}
\put(25,35){\line(-1,0){25}}
\put(25,35){\line(0,-1){35}}
\put(5,13){\hbox{$\mathcal{B}^{(2)}_1$}}
\qbezier(3,0)(3,-8)(6,-8)
\qbezier(9,0)(9,-8)(6,-8)
\qbezier(16,0)(16,-8)(19,-8)
\qbezier(22,0)(22,-8)(19,-8)
}
\put(36,76){
\put(0,0){\line(1,0){25}}
\put(0,0){\line(0,1){35}}
\put(25,35){\line(-1,0){25}}
\put(25,35){\line(0,-1){35}}
\put(5,13){\hbox{$\mathcal{B}^{(2)}_n$}}
\qbezier(3,35)(3,43)(6,43)
\qbezier(9,35)(9,43)(6,43)
\qbezier(16,35)(16,43)(19,43)
\qbezier(22,35)(22,43)(19,43)
}

\put(9,-15){\line(0,1){28}}
\put(3,-15){\line(0,1){28}}
\put(57,-15){\line(0,1){28}}
\put(51,-15){\line(0,1){28}}
\put(16,-15){\line(0,1){28}}
\put(22,-15){\line(0,1){28}}
\put(39,-15){\line(0,1){28}}
\put(45,-15){\line(0,1){28}}
\put(9,48){\line(0,1){28}}
\put(3,48){\line(0,1){28}}
\put(57,48){\line(0,1){28}}
\put(51,48){\line(0,1){28}}
\put(16,48){\line(0,1){28}}
\put(22,48){\line(0,1){28}}
\put(39,48){\line(0,1){28}}
\put(45,48){\line(0,1){28}}
\put(0,13){
\put(0,0){\line(1,0){25}}
\put(0,0){\line(0,1){7}}
\put(25,0){\line(0,1){7}}
\put(25,35){\line(-1,0){25}}
\put(25,35){\line(0,-1){7}}
\put(0,35){\line(0,-1){7}}
\put(5,9){\hbox{$\,\cdot\cdot\cdot$}}
\put(5,20){\hbox{$\,\cdot\cdot\cdot$}}
}
\put(36,13){
\put(0,0){\line(1,0){25}}
\put(0,0){\line(0,1){7}}
\put(25,0){\line(0,1){7}}
\put(25,35){\line(-1,0){25}}
\put(25,35){\line(0,-1){7}}
\put(0,35){\line(0,-1){7}}
\put(5,9){\hbox{$\,\cdot\cdot\cdot$}}
\put(5,20){\hbox{$\,\cdot\cdot\cdot$}}
}
\end{picture}
\caption{Two unentangled qubits
\label{2free}}
\end{subfigure}%
 \begin{subfigure}{.35\textwidth}
\begin{picture}(200,190)(-50,-75)
\put(0,-50){
\put(0,0){\line(1,0){25}}
\put(0,0){\line(0,1){35}}
\put(25,35){\line(-1,0){25}}
\put(25,35){\line(0,-1){35}}
\put(5,13){\hbox{$\mathcal{B}^{(1)}_{1}$}}
\qbezier(3,0)(3,-8)(6,-8)
\qbezier(9,0)(9,-8)(6,-8)
\qbezier(16,0)(16,-8)(19,-8)
\qbezier(22,0)(22,-8)(19,-8)
}
\put(0,75){
\put(0,0){\line(1,0){25}}
\put(0,0){\line(0,1){35}}
\put(25,35){\line(-1,0){25}}
\put(25,35){\line(0,-1){35}}
\put(5,13){\hbox{$\mathcal{B}^{(1)}_{2}$}}
\qbezier(3,35)(3,43)(6,43)
\qbezier(9,35)(9,43)(6,43)
\qbezier(16,35)(16,43)(19,43)
\qbezier(22,35)(22,43)(19,43)
}
\put(32,-50){
\put(0,0){\line(1,0){25}}
\put(0,0){\line(0,1){35}}
\put(25,35){\line(-1,0){25}}
\put(25,35){\line(0,-1){35}}
\put(5,13){\hbox{$\mathcal{B}^{(2)}_1$}}
\qbezier(3,0)(3,-8)(6,-8)
\qbezier(9,0)(9,-8)(6,-8)
\qbezier(16,0)(16,-8)(19,-8)
\qbezier(22,0)(22,-8)(19,-8)
}
\put(32,75){
\put(0,0){\line(1,0){25}}
\put(0,0){\line(0,1){35}}
\put(25,35){\line(-1,0){25}}
\put(25,35){\line(0,-1){35}}
\put(5,13){\hbox{$\mathcal{B}^{(2)}_2$}}
\qbezier(3,35)(3,43)(6,43)
\qbezier(9,35)(9,43)(6,43)
\qbezier(16,35)(16,43)(19,43)
\qbezier(22,35)(22,43)(19,43)
}

\put(1.5,-15){\line(0,1){5}}
\put(7.5,-15){\line(0,1){5}}

\put(1.5,-10){\qbezier(0,0)(0,2.5)(3,5)\qbezier(3,5)(6,7.5)(6,10)\qbezier(6,0)(6,2.5)(4,4)\qbezier(2,6)(0,7,5)(0,10)}
\put(1.5,0){\qbezier(0,0)(0,2.5)(3,5)\qbezier(3,5)(6,7.5)(6,10)\qbezier(6,0)(6,2.5)(4,4)\qbezier(2,6)(0,7,5)(0,10)}

\put(1.5,10){\line(0,1){5}}
\put(7.5,10){\line(0,1){5}}

\put(1.5,45){\qbezier(0,0)(0,2.5)(3,5)\qbezier(3,5)(6,7.5)(6,10)\qbezier(6,0)(6,2.5)(4,4)\qbezier(2,6)(0,7,5)(0,10)}
\put(1.5,55){\qbezier(0,0)(0,2.5)(3,5)\qbezier(3,5)(6,7.5)(6,10)\qbezier(6,0)(6,2.5)(4,4)\qbezier(2,6)(0,7,5)(0,10)}
\put(1.5,65){\qbezier(0,0)(0,2.5)(3,5)\qbezier(3,5)(6,7.5)(6,10)\qbezier(6,0)(6,2.5)(4,4)\qbezier(2,6)(0,7,5)(0,10)}
\put(55.5,-15){\line(0,1){10}}
\put(49.5,-15){\line(0,1){10}}
\put(55.5,5){\line(0,1){40}}
\put(49.5,5){\line(0,1){10}}
\put(39.5,15){\qbezier(0,0)(0,4)(5,7)\qbezier(5,7)(10,10)(10,14)\qbezier(10,0)(10,4)(6,6)\qbezier(4,8)(0,10)(0,14)}
\put(39.5,29){\qbezier(0,0)(0,4)(5,7)\qbezier(5,7)(10,10)(10,14)\qbezier(10,0)(10,4)(6,6)\qbezier(4,8)(0,10)(0,14)}
\put(49.5,43){\line(0,1){2}}
\put(49.5,-5){\qbezier(0,0)(0,2.5)(3,5)\qbezier(3,5)(6,7.5)(6,10)\qbezier(6,0)(6,2.5)(4,4)\qbezier(2,6)(0,7,5)(0,10)}
\put(49.5,45){\qbezier(0,0)(0,2.5)(3,5)\qbezier(3,5)(6,7.5)(6,10)\qbezier(6,0)(6,2.5)(4,4)\qbezier(2,6)(0,7,5)(0,10)}
\put(49.5,55){\qbezier(0,0)(0,2.5)(3,5)\qbezier(3,5)(6,7.5)(6,10)\qbezier(6,0)(6,2.5)(4,4)\qbezier(2,6)(0,7,5)(0,10)}
\put(49.5,65){\qbezier(0,0)(0,2.5)(3,5)\qbezier(3,5)(6,7.5)(6,10)\qbezier(6,0)(6,2.5)(4,4)\qbezier(2,6)(0,7,5)(0,10)}
\put(33.5,15){\line(0,1){30}}\put(39.5,43){\line(0,1){2}}

\put(17.5,-15){\qbezier(0,0)(0,8)(8,15.5)\qbezier(8,15.5)(16,23)(16,30)\qbezier(16,0)(16,8)(12.5,10.5)\qbezier(6,16)(0,23)(0,30)
\qbezier(6,0)(6,8)(14.5,15)\qbezier(14.5,15)(22,23)(22,30)\qbezier(22,0)(22,8)(16,13.5)\qbezier(9.5,19)(6,23)(6,30)}

\put(1.5,15){\qbezier(22,0)(22,8)(16,13.5)\qbezier(13,16)(6,23)(6,30)\qbezier(6,0)(6,8)(12,13)\qbezier(12,13)(22,23)(22,30)
\qbezier(16,0)(16,8)(12.5,11)\qbezier(9,14)(0,23)(0,30)\qbezier(0,0)(0,9)(6,14)\qbezier(12,19.5)(16,23)(16,30)}

\put(17.5,45){\qbezier(0,0)(0,8)(6,13.5)\qbezier(8.5,16)(16,23)(16,30)\qbezier(16,0)(16,8)(12.5,10.5)\qbezier(9.5,12.5)(0,23)(0,30)
\qbezier(6,0)(6,8)(13.5,14)\qbezier(15.5,16)(22,23)(22,30)\qbezier(22,0)(22,8)(12,17.5)\qbezier(9.5,19)(6,23)(6,30)}

\end{picture}
\caption{Two qubits with entangled strands.
\label{2ent0}}
\end{subfigure}%
 \begin{subfigure}{.35\textwidth}
\begin{picture}(200,190)(-50,-75)
\put(0,-50){
\put(0,0){\line(1,0){25}}
\put(0,0){\line(0,1){35}}
\put(25,35){\line(-1,0){25}}
\put(25,35){\line(0,-1){35}}
\put(5,13){\hbox{$\mathcal{B}^{(1)}_{1}$}}
\qbezier(3,0)(3,-8)(6,-8)
\qbezier(9,0)(9,-8)(6,-8)
\qbezier(16,0)(16,-8)(19,-8)
\qbezier(22,0)(22,-8)(19,-8)
}
\put(0,76){
\put(0,0){\line(1,0){25}}
\put(0,0){\line(0,1){35}}
\put(25,35){\line(-1,0){25}}
\put(25,35){\line(0,-1){35}}
\put(5,13){\hbox{$\mathcal{B}^{(1)}_{2}$}}
\qbezier(3,35)(3,43)(6,43)
\qbezier(9,35)(9,43)(6,43)
\qbezier(16,35)(16,43)(19,43)
\qbezier(22,35)(22,43)(19,43)
}
\put(36,-50){
\put(0,0){\line(1,0){25}}
\put(0,0){\line(0,1){35}}
\put(25,35){\line(-1,0){25}}
\put(25,35){\line(0,-1){35}}
\put(5,13){\hbox{$\mathcal{B}^{(2)}_1$}}
\qbezier(3,0)(3,-8)(6,-8)
\qbezier(9,0)(9,-8)(6,-8)
\qbezier(16,0)(16,-8)(19,-8)
\qbezier(22,0)(22,-8)(19,-8)
}
\put(36,76){
\put(0,0){\line(1,0){25}}
\put(0,0){\line(0,1){35}}
\put(25,35){\line(-1,0){25}}
\put(25,35){\line(0,-1){35}}
\put(5,13){\hbox{$\mathcal{B}^{(2)}_2$}}
\qbezier(3,35)(3,43)(6,43)
\qbezier(9,35)(9,43)(6,43)
\qbezier(16,35)(16,43)(19,43)
\qbezier(22,35)(22,43)(19,43)
}

\put(9,-15){\line(0,1){91}}
\put(3,-15){\line(0,1){91}}
\put(57,-15){\line(0,1){91}}
\put(51,-15){\line(0,1){91}}
\put(16,-15){\line(0,1){20}}
\put(22,-15){\line(0,1){20}}
\put(39,-15){\line(0,1){20}}
\put(45,-15){\line(0,1){20}}
\put(16,56){\line(0,1){20}}
\put(22,56){\line(0,1){20}}
\put(39,56){\line(0,1){20}}
\put(45,56){\line(0,1){20}}

\put(16,5){\qbezier(0,0)(0,13)(10,18)
\qbezier(16,21)(25,25)(15,30)
\qbezier(15,30)(0,35)(0,51)}
\put(22,5){\qbezier(0,0)(0,12)(8,15)
\qbezier(14,19)(25,25)(15,30)
\qbezier(15,30)(0,35)(0,51)}
\put(39,5){\qbezier(0,0)(0,15)(-15,20)
\qbezier(-15,20)(-25,25)(-14,31)
\qbezier(-8,35)(0,38)(0,51)}
\put(45,5){\qbezier(0,0)(0,15)(-15,20)
\qbezier(-15,20)(-25,25)(-16,29)
\qbezier(-10,32)(0,37)(0,51)}

\end{picture}
\caption{Two qubits entangled with cables.
\label{2ent1}}
\end{subfigure}
\caption{Operation, which entangles two qubits, can be constructed by braiding cables of two strands to form a more complex knot. If we entangle them by braiding individual strands, the Hilbert space of the structure will be much larger than the computational space of two qubits.}
\end{figure}
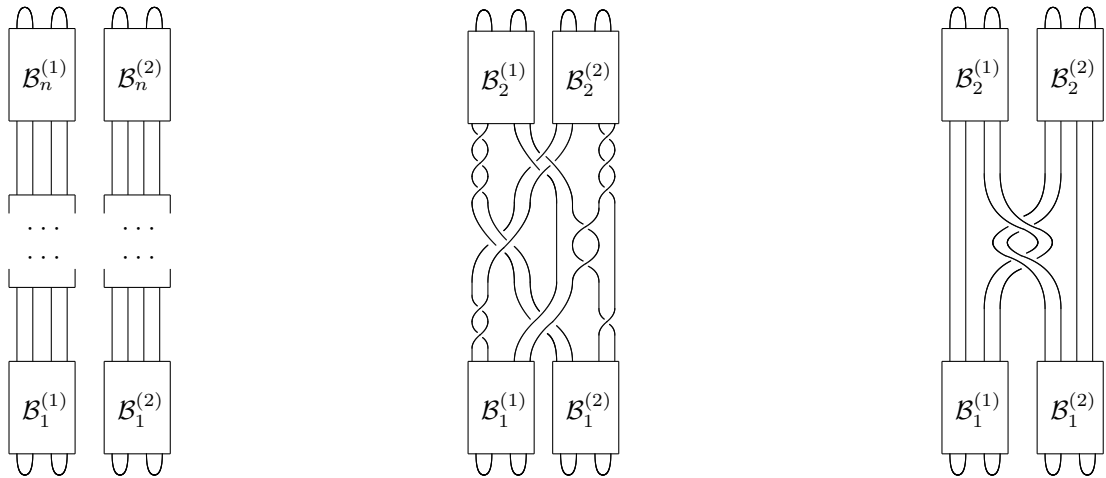
\begin{figure}[h!]
 \begin{subfigure}{.4\textwidth}
\begin{picture}(200,270)(-75,-75)
\put(0,-50){
\put(0,0){\line(1,0){25}}
\put(0,0){\line(0,1){35}}
\put(25,35){\line(-1,0){25}}
\put(25,35){\line(0,-1){35}}
\put(5,13){\hbox{$\mathcal{B}^{(1)}_{1}$}}
\put(3,0){\line(0,-1){5}}
\put(3,-8){\line(0,-1){3}}
\put(9,0){\line(0,-1){5}}
\put(9,-8){\line(0,-1){3}}
\put(16,0){\line(0,-1){5}}
\put(16,-8){\line(0,-1){3}}
\put(22,0){\line(0,-1){5}}
\put(22,-8){\line(0,-1){3}}
\put(3,-14){\line(0,-1){3}}
\put(9,-14){\line(0,-1){3}}
\put(16,-14){\line(0,-1){3}}
\put(22,-14){\line(0,-1){3}}
}
\put(0,105){
\put(0,0){\line(1,0){25}}
\put(0,0){\line(0,1){35}}
\put(25,35){\line(-1,0){25}}
\put(25,35){\line(0,-1){35}}
\put(5,13){\hbox{$\mathcal{B}^{(1)}_{2}$}}
\put(3,35){\line(0,1){5}}
\put(3,43){\line(0,1){3}}
\put(9,35){\line(0,1){5}}
\put(9,43){\line(0,1){3}}
\put(16,35){\line(0,1){5}}
\put(16,43){\line(0,1){3}}
\put(22,35){\line(0,1){5}}
\put(22,43){\line(0,1){3}}
\put(3,49){\line(0,1){3}}
\put(9,49){\line(0,1){3}}
\put(16,49){\line(0,1){3}}
\put(22,49){\line(0,1){3}}
}
\put(32,-50){
\put(0,0){\line(1,0){25}}
\put(0,0){\line(0,1){35}}
\put(25,35){\line(-1,0){25}}
\put(25,35){\line(0,-1){35}}
\put(5,13){\hbox{$\mathcal{B}^{(2)}_1$}}
\put(3,0){\line(0,-1){5}}
\put(3,-8){\line(0,-1){3}}
\put(9,0){\line(0,-1){5}}
\put(9,-8){\line(0,-1){3}}
\put(16,0){\line(0,-1){5}}
\put(16,-8){\line(0,-1){3}}
\put(22,0){\line(0,-1){5}}
\put(22,-8){\line(0,-1){3}}
\put(3,-14){\line(0,-1){3}}
\put(9,-14){\line(0,-1){3}}
\put(16,-14){\line(0,-1){3}}
\put(22,-14){\line(0,-1){3}}
}
\put(32,105){
\put(0,0){\line(1,0){25}}
\put(0,0){\line(0,1){35}}
\put(25,35){\line(-1,0){25}}
\put(25,35){\line(0,-1){35}}
\put(5,13){\hbox{$\mathcal{B}^{(2)}_2$}}
\put(3,35){\line(0,1){5}}
\put(3,43){\line(0,1){3}}
\put(9,35){\line(0,1){5}}
\put(9,43){\line(0,1){3}}
\put(16,35){\line(0,1){5}}
\put(16,43){\line(0,1){3}}
\put(22,35){\line(0,1){5}}
\put(22,43){\line(0,1){3}}
\put(3,49){\line(0,1){3}}
\put(9,49){\line(0,1){3}}
\put(16,49){\line(0,1){3}}
\put(22,49){\line(0,1){3}}
}

\green\put(1.5,-15){\line(0,1){30}}
\put(1.5,75){\line(0,1){30}}
\put(7.5,-15){\line(0,1){30}}
\put(7.5,75){\line(0,1){30}}
\black
\put(55.5,-15){\line(0,1){120}}
\put(49.5,-15){\line(0,1){120}}

{\red\put(33.5,15){\line(0,1){60}}\put(39.5,15){\line(0,1){60}}}

\put(17.5,-15){\red\qbezier(0,0)(0,8)(8,15.5)\qbezier(8,15.5)(16,23)(16,30)\blue\qbezier(16,0)(16,8)(12.5,10.5)\qbezier(6,16)(0,23)(0,30)
\red\qbezier(6,0)(6,8)(14.5,15)\qbezier(14.5,15)(22,23)(22,30)\blue\qbezier(22,0)(22,8)(16,13.5)\qbezier(9.5,19)(6,23)(6,30)}

\put(1.5,15){\blue\qbezier(22,0)(22,8)(14,15.5)\qbezier(14,15.5)(6,23)(6,30)\green\qbezier(6,0)(6,8)(9.5,10.5)\qbezier(16,16)(22,23)(22,30)
\blue\qbezier(16,0)(16,8)(7.5,15)\qbezier(7.5,15)(0,23)(0,30)\green\qbezier(0,0)(0,8)(6,13.5)\qbezier(12.5,19)(16,23)(16,30)}

\put(1.5,45){\green\qbezier(22,0)(22,8)(14,15.5)\qbezier(14,15.5)(6,23)(6,30)\blue\qbezier(6,0)(6,8)(9.5,10.5)\qbezier(16,16)(22,23)(22,30)
\green\qbezier(16,0)(16,8)(7.5,15)\qbezier(7.5,15)(0,23)(0,30)\blue\qbezier(0,0)(0,8)(6,13.5)\qbezier(12.5,19)(16,23)(16,30)}

\put(17.5,75){\blue\qbezier(0,0)(0,8)(8,15.5)\qbezier(8,15.5)(16,23)(16,30)\red\qbezier(16,0)(16,8)(12.5,10.5)\qbezier(6,16)(0,23)(0,30)
\blue\qbezier(6,0)(6,8)(14.5,15)\qbezier(14.5,15)(22,23)(22,30)\red\qbezier(22,0)(22,8)(16,13.5)\qbezier(9.5,19)(6,23)(6,30)}

\put(-20,-20){\line(1,0){100}}
\put(-20,110){\line(1,0){100}}
\put(-20,-20){\line(0,1){130}}
\put(80,-20){\line(0,1){130}}
\end{picture}
\end{subfigure}
\begin{subfigure}{.1\textwidth}
\begin{picture}(200,270)(0,-75)

\put(1.5,50){\line(1,0){30}}
\put(31.5,50){\line(-1,1){10}}
\put(31.5,50){\line(-1,-1){10}}
\end{picture}
\end{subfigure}
\begin{subfigure}{.5\textwidth}
\begin{picture}(200,270)(-75,-75)

\green\put(1.5,-15){\line(0,1){30}}
\put(1.5,75){\line(0,1){30}}
\black\put(50.5,-15){\line(0,1){120}}

\red\put(33.5,15){\line(0,1){60}}

\put(17.5,-15){\red\qbezier(0,0)(0,8)(8,15.5)\qbezier(8,15.5)(16,23)(16,30)\blue\qbezier(16,0)(16,8)(10.5,12.5)\qbezier(5,17)(0,23)(0,30)}

\put(1.5,15){\blue\qbezier(16,0)(16,8)(8.5,15)\qbezier(8.5,15)(0,23)(0,30)\green\qbezier(0,0)(0,8)(5.5,13)\qbezier(10.5,17)(16,23)(16,30)}

\put(1.5,45){\green\qbezier(16,0)(16,8)(8.5,15)\qbezier(8.5,15)(0,23)(0,30)\blue\qbezier(0,0)(0,8)(5.5,13)\qbezier(10.5,17)(16,23)(16,30)}

\put(17.5,75){\blue\qbezier(0,0)(0,8)(8,15.5)\qbezier(8,15.5)(16,23)(16,30)\red\qbezier(16,0)(16,8)(10.5,12.5)\qbezier(5,17)(0,23)(0,30)}
\put(0.4,0){\green\put(1.5,-15){\line(0,1){30}}
\put(1.5,75){\line(0,1){30}}
\black\put(50.5,-15){\line(0,1){120}}

\red\put(33.5,15){\line(0,1){60}}

\put(17.5,-15){\red\qbezier(0,0)(0,8)(8,15.5)\qbezier(8,15.5)(16,23)(16,30)\blue\qbezier(16,0)(16,8)(10.5,12.5)\qbezier(5,17)(0,23)(0,30)}

\put(1.5,15){\blue\qbezier(16,0)(16,8)(8.5,15)\qbezier(8.5,15)(0,23)(0,30)\green\qbezier(0,0)(0,8)(5.5,13)\qbezier(10.5,17)(16,23)(16,30)}

\put(1.5,45){\green\qbezier(16,0)(16,8)(8.5,15)\qbezier(8.5,15)(0,23)(0,30)\blue\qbezier(0,0)(0,8)(5.5,13)\qbezier(10.5,17)(16,23)(16,30)}

\put(17.5,75){\blue\qbezier(0,0)(0,8)(8,15.5)\qbezier(8,15.5)(16,23)(16,30)\red\qbezier(16,0)(16,8)(10.5,12.5)\qbezier(5,17)(0,23)(0,30)}
}
\end{picture}
\end{subfigure}
\caption{Entangling of two qubits with the cables on the left can be interpreted as a braid in higher representations on the right. These strands can be either in trivial or adjoint representation, leading to the different cases from (\ref{cases}). \label{2ent3}}
\end{figure}
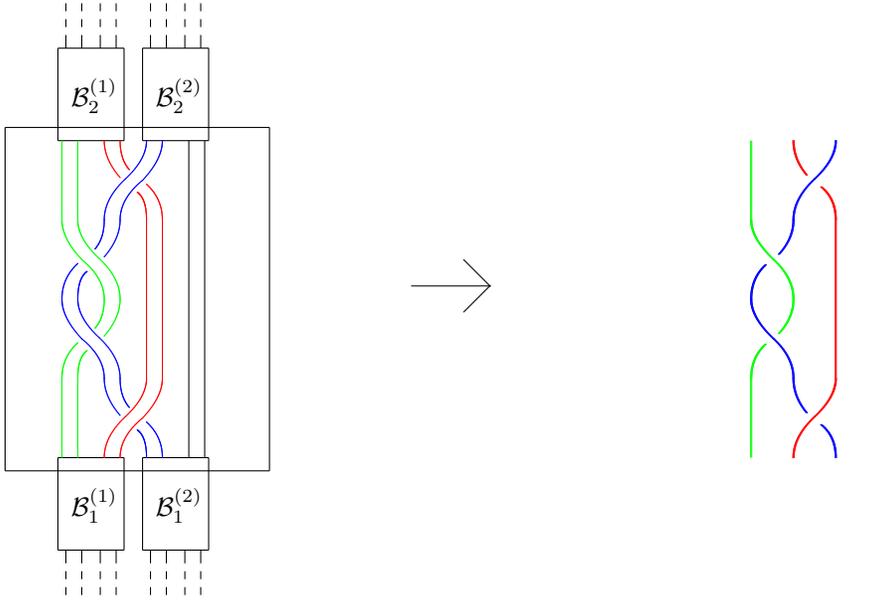

However, the situation is more complex when we consider general types of intersections.

In full analogy to the case of a four strands diagrams, now we should take a product of representations on all eight strands and then take all trivial representations, which appear there. In the case of $U_q(SU(N))$ group we take the following product:
\begin{equation}
([1]\otimes\bar{[1]})\otimes([1]\otimes\bar{[1]})\otimes([1]\otimes\bar{[1]})\otimes([1]\otimes\bar{[1]})=24\ \emptyset +\ldots
\label{8prod}
\end{equation}
In the case of $U_q(SU(2))$ this formula reduces\footnote{In the case of $U_q(SU(3))$ representation $[1,1,1,1]$ is absent, hence the product in consideration results in 23 trivial representation} (moreover, $\bar{[1]}=[1]$ in this case):
\begin{equation}
([1]\otimes{[1]})\otimes([1]\otimes{[1]})\otimes([1]\otimes{[1]})\otimes([1]\otimes{[1]})=14\ \emptyset +\ldots
\end{equation}
Therefore, if we put two one-qubit next to each other and start entangling their strands in an arbitrary way, see Fig.\ref{2ent0}, we will move from the four-dimensional space to the 24-dimensional from which only the four-dimensional space is computable (corresponds to the space of qubits). However, by reducing the arbitrariness of the entanglement we can greatly reduce the probability of moving away from the computational space.

First simplification is that instead of entangling all the strands, we should entangle cables made of two strands (see Fig.\ref{2ent1}). This in principle can be interpreted as strands in higher representations \cite{Cable}.

Therefore instead of the product (\ref{8prod}) we should do the following. Firstly, in each qubit - each 4-plat - we take two pairs of strands which can be either in trivial or in adjoint representation:
\begin{equation}
[1]\otimes\bar{[1]}=\emptyset+\text{adj}.
\end{equation}
Representation, running in the cable, is not changed if we do not separate strands in the cable and do not entangle them separately with other strands. Therefore we can just consider the representations in four cables. This means that due to the (\ref{1q}) and initial states of qubits, the space of representations reduces to the following four cases\footnote{In this classification we already assumed that after intertwining the cables do not change the position, i.e. they intersect even number of times and return to the initial position, as in Fig.\ref{2ent3} for example.}:
\begin{equation}
\begin{array}{lcl}
I.&\ &\emptyset \otimes \emptyset \otimes \emptyset \otimes \emptyset,
\\
II.&\ &\emptyset \otimes \emptyset \otimes \text{adj} \otimes \text{adj},
\\
III.&\ &\text{adj} \otimes \text{adj} \otimes \emptyset \otimes \emptyset,
\\
IV.&\ &\text{adj} \otimes \text{adj} \otimes \text{adj} \otimes \text{adj}.
\end{array}
\label{cases}
\end{equation}
There is no case which includes $\emptyset\otimes \text{adj}$ on one side since we want to start from the computable space where both sides are in trivial representation, which cannot be achieved if there is a product $\emptyset\otimes \text{adj}$.

After this type of entangling, (see for example Fig.\ref{2ent3}), representations transform in the following way. Trivial representation $\emptyset$ interacts trivially with all other representations. Therefore, case $I.$ will not be changed by any braiding.
Cases $II.$ and $III.$ also do not change as long as cables do not change the position, i.e. one keeps the left pair on the left in the end of the intertwining and the right pair on the right\footnote{It is possible also to move the right pair as a whole to the left, and the left pair to the right, but this is more difficult to understand and study the corresponding operator. However, if we separate two cables and move only one of them to the other side, the whole structure will be destroyed.}. Since two of the cables on one side are in trivial representation and do not interact with other cables, there is basically no two-qubit interaction. The whole braid can be reduced to the one-qubit interactions on the side with two adjoint representations (right side in case $II.$ or left side in case $III.$).

The non-trivial is case $IV.$ Here each cable is in the adjoint representation. It is easy to see that the whole picture becomes similar to the one-qubit state in section \ref{s:1qb}, but instead of fundamental and anti-fundamental representations, each strand carries adjoint representation (adjoint representation is self-conjugate therefore anti-adjoint is also an adjoint representation), see Figure \ref{2ent3} for an example. It is known \cite{UniRac} that in the product $\text{adj}\otimes\text{adj}\otimes\text{adj}\otimes\text{adj}$ there are 9 trivial representations\footnote{9 is for the case of $U_q(SU(N))$, $N\ge4$. It would be 8 trivial representations for the $U_q(SU(3))$ case and 3 for $U_q(SU(2))$.}, rather than two in the case of fundamental and anti-fundamental representation. For the case of $U_q(SU(2))$, this reduces to the $[2]\otimes [2]\otimes [2]\otimes [2]$ where there are only three trivial representations. However, in both cases, only one of these trivial representations correspond to our computable space, where each of the qubits is in the trivial representation.

Therefore to have an efficient two-qubit operation, we need a non-trivial braiding which in the case $IV.$ keeps the system in computational space. In other words, upon two-qubit intertwining, the probability of the system in the case $IV.$ to leave the computable space should be vanishing, while the phase shift, given by such a braiding, should be nontrivial. In the next section we will discuss how this works in the simplest $U_q(SU(2))$ case and present an explicit examples of such operators.

Since interactions for all the cases, but $IV.$ is trivial, the corresponding two-qubit operator will have the following form
\begin{equation}
\mathcal{O}\approx\left(
  \begin{array}{cccc}
  1
  \\
  & 1\\
  && 1\\
  &&& e^{i\phi}
  \end{array}
\right),
\end{equation}
where $\phi$ is a corresponding phase shift, which depends on a particular set of intersections and parameters $q$ and $N$.

\section{$U_q(SU(2))$ braiding\label{s:2braid}}

In this section we consider the case of $U_q(SU(2))$ and provide the explicit construction. We use the following formulae for braiding and mixing matrices\footnote{Three representations that constitute the space where these two matrices act are $\emptyset$, $[2]$ (or {\it adj}) and $[4]$} \cite{LargeK,NPZ,6jsym,6jsym2}:
\begin{equation}
R=\left(\begin{array}{ccc}
1
\\
& -q^2
\\
&& q^6
\end{array}\right)
\label{R2r}
\end{equation}
and
\begin{equation}
S=\bar{S}=\left(\begin{array}{ccc}
\cfrac{1}{[3]_q} & \cfrac{\sqrt{[3]_q}}{[3]_q} & \cfrac{\sqrt{[5]_q}}{[3]_q}
\\
\cfrac{\sqrt{[3]_q}}{[3]_q} & \cfrac{[6]_q}{[3]_q[4]_q} & -\cfrac{[2]_q\sqrt{[3]_q[5]_q}}{[3]_q[4]_q}
\\
-\cfrac{\sqrt{[5]_q}}{[3]_q} & \cfrac{[2]_q\sqrt{[3]_q[5]_q}}{[3]_q[4]_q} & -\cfrac{[2]_q}{[3]_q[4]_q}
\end{array}\right).
\label{U2r}
\end{equation}

\begin{figure}[h!]
\begin{picture}(200,100)(-175,-75)
\put(-50,-50){
\put(0,0){\line(0,1){10}}
\qbezier(0,10)(0,22)(12,34)
\qbezier(36,46)(24,46)(12,34)
\qbezier(36,46)(48,46)(48,34)
\qbezier(36,22)(48,22)(48,34)
\qbezier(36,22)(24,22)(14,32)
\qbezier(10,36)(0,46)(0,58)
\put(0,58){\line(0,1){10}}
\put(60,32){\hbox{$=$}}
\put(80,0){\line(0,1){68}}
\put(35,-15){\hbox{$T$}}
}
\put(150,-50){
\put(24,58){\line(0,1){10}}
\qbezier(0,10)(0,22)(12,34)
\qbezier(24,58)(24,46)(12,34)
\qbezier(0,58)(0,46)(10,36)
\qbezier(14,32)(24,22)(24,10)
\put(0,58){\line(0,1){10}}
\qbezier(24,10)(24,-2)(12,-2)
\qbezier(0,10)(0,-2)(12,-2)
\put(50,32){\hbox{$=$}}
\put(80,10){\line(0,1){58}}
\put(104,10){\line(0,1){58}}
\qbezier(104,10)(104,-2)(92,-2)
\qbezier(80,10)(80,-2)(92,-2)
\put(55,-15){\hbox{$\bar{T}$}}
}
\end{picture}
\caption{Unknotting property of $T$ and $\bar{T}$ operators.
\label{Tmat}}
\end{figure}
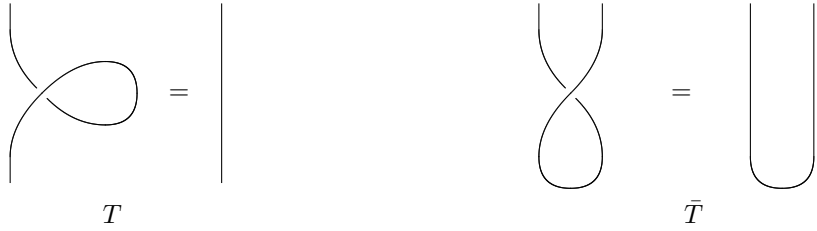

Due to the properties of the $U_q(SU(2))$ algebra, all the representations are self-conjugate. Therefore, the matrices $S$ and $\bar{S}$ coincide. Similarly matrices $T$ and $\bar{T}$ should also coincide, but up to a general coefficient. This is due to the different additional property of these matrices. There are in fact three different braiding matrices.
Depending on a set of intersections, 4-plat picture can correspond either to a knot or to a link. In the latter case there are two separate link components, each with its own representation. While the representation and its conjugate are equivalent in the $U_q(SU(2))$ algebra, they are not exactly equal in the same basis. This means that the system should not be able to distinguish them when the strands correspond to two different components, because the basis is not fixed between the different components. We will denote this crossing by $T_{i}$, and the operator is given by the formulae in \cite{Klimyk}. However, the matrices that correspond to the intersection of the one same component can depend on the direction of the strands, or, in other words they can distinguish representation and its conjugate. These matrices are defined by the properties on Fig.\ref{Tmat} and are equal to:

\begin{equation}
\bar{T}=R,\ \ \ T=q^{-4}R,\ \ \ T_i=q^{-8}R.
\end{equation}

Using these matrices one can construct an operator $\mathcal{B}$ that corresponds to the braiding of the adjoint strands. We are interested in the matrix element $\mathcal{B}_{1,1}$ of this matrix, which corresponds to the transformation of two-qubits that are initially in a computational space (case $IV.$ from (\ref{cases})) and are kept there in the end. Therefore, probability of two qubits to remain in computable space $\mathcal{P}$ and phase shift $\phi$ are given by
\begin{equation}
\mathcal{P}=|\mathcal{B}_{1,1}|^2,\ \ \ \phi=\text{Arg}(\mathcal{B}_{1,1}).
\end{equation}

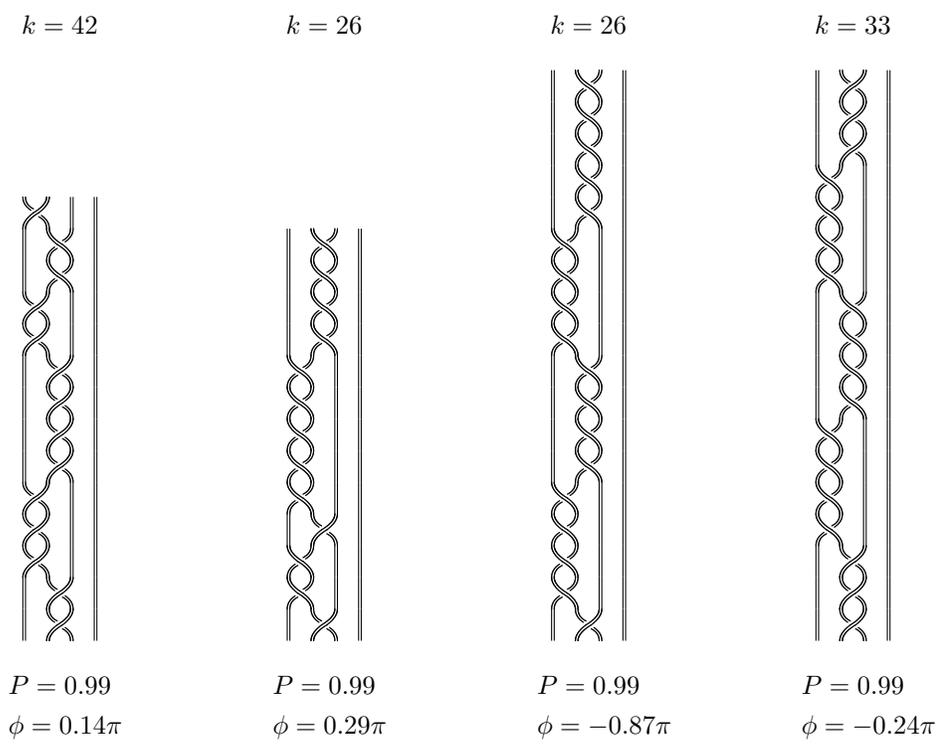
\begin{figure}[h!]
\begin{picture}(200,300)(-175,-40)

\put(-110,230){\hbox{$k=42$}}
\put(-115,-20){\hbox{$P=0.99$}}
\put(-115,-35){\hbox{$\phi=0.14\pi$}}
\put(-100,0){\qbezier(0.5,0)(0.5,3)(5.5,6)\qbezier(5.5,6)(9.5,9)(9.5,12)\qbezier(-0.5,0)(-0.5,3)(3.5,6)\qbezier(3.5,6)(8.5,9)(8.5,12)
\qbezier(9.5,0)(9.5,3)(6.5,5)\qbezier(3.5,7.5)(0.5,9)(0.5,12)\qbezier(8.5,0)(8.5,3)(5.5,4.5)\qbezier(2.5,7)(-0.5,9)(-0.5,12)
\put(18,0){\put(-0.5,0){\line(0,1){12}}\put(0.5,0){\line(0,1){12}}}\put(-18,0){\put(8.5,0){\line(0,1){12}}\put(9.5,0){\line(0,1){12}}}
}

\put(-100,12){\qbezier(0.5,0)(0.5,3)(5.5,6)\qbezier(5.5,6)(9.5,9)(9.5,12)\qbezier(-0.5,0)(-0.5,3)(3.5,6)\qbezier(3.5,6)(8.5,9)(8.5,12)
\qbezier(9.5,0)(9.5,3)(6.5,5)\qbezier(3.5,7.5)(0.5,9)(0.5,12)\qbezier(8.5,0)(8.5,3)(5.5,4.5)\qbezier(2.5,7)(-0.5,9)(-0.5,12)
\put(18,0){\put(-0.5,0){\line(0,1){12}}\put(0.5,0){\line(0,1){12}}}\put(-18,0){\put(8.5,0){\line(0,1){12}}\put(9.5,0){\line(0,1){12}}}
}

\put(-100,24){
\put(-9,0){\qbezier(0.5,0)(0.5,3)(5.5,6)\qbezier(5.5,6)(9.5,9)(9.5,12)\qbezier(-0.5,0)(-0.5,3)(3.5,6)\qbezier(3.5,6)(8.5,9)(8.5,12)
\qbezier(9.5,0)(9.5,3)(6.5,5)\qbezier(3.5,7.5)(0.5,9)(0.5,12)\qbezier(8.5,0)(8.5,3)(5.5,4.5)\qbezier(2.5,7)(-0.5,9)(-0.5,12)}
\put(9,0){\put(-0.5,0){\line(0,1){12}}\put(0.5,0){\line(0,1){12}}}\put(9,0){\put(8.5,0){\line(0,1){12}}\put(9.5,0){\line(0,1){12}}}
}

\put(-100,36){
\put(-9,0){\qbezier(0.5,0)(0.5,3)(5.5,6)\qbezier(5.5,6)(9.5,9)(9.5,12)\qbezier(-0.5,0)(-0.5,3)(3.5,6)\qbezier(3.5,6)(8.5,9)(8.5,12)
\qbezier(9.5,0)(9.5,3)(6.5,5)\qbezier(3.5,7.5)(0.5,9)(0.5,12)\qbezier(8.5,0)(8.5,3)(5.5,4.5)\qbezier(2.5,7)(-0.5,9)(-0.5,12)}
\put(9,0){\put(-0.5,0){\line(0,1){12}}\put(0.5,0){\line(0,1){12}}}\put(9,0){\put(8.5,0){\line(0,1){12}}\put(9.5,0){\line(0,1){12}}}
}

\put(-100,48){
\put(-9,0){\qbezier(0.5,0)(0.5,3)(5.5,6)\qbezier(5.5,6)(9.5,9)(9.5,12)\qbezier(-0.5,0)(-0.5,3)(3.5,6)\qbezier(3.5,6)(8.5,9)(8.5,12)
\qbezier(9.5,0)(9.5,3)(6.5,5)\qbezier(3.5,7.5)(0.5,9)(0.5,12)\qbezier(8.5,0)(8.5,3)(5.5,4.5)\qbezier(2.5,7)(-0.5,9)(-0.5,12)}
\put(9,0){\put(-0.5,0){\line(0,1){12}}\put(0.5,0){\line(0,1){12}}}\put(9,0){\put(8.5,0){\line(0,1){12}}\put(9.5,0){\line(0,1){12}}}
}

\put(-100,60){\qbezier(0.5,0)(0.5,3)(5.5,6)\qbezier(5.5,6)(9.5,9)(9.5,12)\qbezier(-0.5,0)(-0.5,3)(3.5,6)\qbezier(3.5,6)(8.5,9)(8.5,12)
\qbezier(9.5,0)(9.5,3)(6.5,5)\qbezier(3.5,7.5)(0.5,9)(0.5,12)\qbezier(8.5,0)(8.5,3)(5.5,4.5)\qbezier(2.5,7)(-0.5,9)(-0.5,12)
\put(18,0){\put(-0.5,0){\line(0,1){12}}\put(0.5,0){\line(0,1){12}}}\put(-18,0){\put(8.5,0){\line(0,1){12}}\put(9.5,0){\line(0,1){12}}}
}

\put(-100,72){\qbezier(0.5,0)(0.5,3)(5.5,6)\qbezier(5.5,6)(9.5,9)(9.5,12)\qbezier(-0.5,0)(-0.5,3)(3.5,6)\qbezier(3.5,6)(8.5,9)(8.5,12)
\qbezier(9.5,0)(9.5,3)(6.5,5)\qbezier(3.5,7.5)(0.5,9)(0.5,12)\qbezier(8.5,0)(8.5,3)(5.5,4.5)\qbezier(2.5,7)(-0.5,9)(-0.5,12)
\put(18,0){\put(-0.5,0){\line(0,1){12}}\put(0.5,0){\line(0,1){12}}}\put(-18,0){\put(8.5,0){\line(0,1){12}}\put(9.5,0){\line(0,1){12}}}
}

\put(-100,84){\qbezier(0.5,0)(0.5,3)(5.5,6)\qbezier(5.5,6)(9.5,9)(9.5,12)\qbezier(-0.5,0)(-0.5,3)(3.5,6)\qbezier(3.5,6)(8.5,9)(8.5,12)
\qbezier(9.5,0)(9.5,3)(6.5,5)\qbezier(3.5,7.5)(0.5,9)(0.5,12)\qbezier(8.5,0)(8.5,3)(5.5,4.5)\qbezier(2.5,7)(-0.5,9)(-0.5,12)
\put(18,0){\put(-0.5,0){\line(0,1){12}}\put(0.5,0){\line(0,1){12}}}\put(-18,0){\put(8.5,0){\line(0,1){12}}\put(9.5,0){\line(0,1){12}}}
}

\put(-100,96){\qbezier(0.5,0)(0.5,3)(5.5,6)\qbezier(5.5,6)(9.5,9)(9.5,12)\qbezier(-0.5,0)(-0.5,3)(3.5,6)\qbezier(3.5,6)(8.5,9)(8.5,12)
\qbezier(9.5,0)(9.5,3)(6.5,5)\qbezier(3.5,7.5)(0.5,9)(0.5,12)\qbezier(8.5,0)(8.5,3)(5.5,4.5)\qbezier(2.5,7)(-0.5,9)(-0.5,12)
\put(18,0){\put(-0.5,0){\line(0,1){12}}\put(0.5,0){\line(0,1){12}}}\put(-18,0){\put(8.5,0){\line(0,1){12}}\put(9.5,0){\line(0,1){12}}}
}

\put(-100,108){
\put(-9,0){\qbezier(0.5,0)(0.5,3)(5.5,6)\qbezier(5.5,6)(9.5,9)(9.5,12)\qbezier(-0.5,0)(-0.5,3)(3.5,6)\qbezier(3.5,6)(8.5,9)(8.5,12)
\qbezier(9.5,0)(9.5,3)(6.5,5)\qbezier(3.5,7.5)(0.5,9)(0.5,12)\qbezier(8.5,0)(8.5,3)(5.5,4.5)\qbezier(2.5,7)(-0.5,9)(-0.5,12)}
\put(9,0){\put(-0.5,0){\line(0,1){12}}\put(0.5,0){\line(0,1){12}}}\put(9,0){\put(8.5,0){\line(0,1){12}}\put(9.5,0){\line(0,1){12}}}
}

\put(-100,120){
\put(-9,0){\qbezier(0.5,0)(0.5,3)(5.5,6)\qbezier(5.5,6)(9.5,9)(9.5,12)\qbezier(-0.5,0)(-0.5,3)(3.5,6)\qbezier(3.5,6)(8.5,9)(8.5,12)
\qbezier(9.5,0)(9.5,3)(6.5,5)\qbezier(3.5,7.5)(0.5,9)(0.5,12)\qbezier(8.5,0)(8.5,3)(5.5,4.5)\qbezier(2.5,7)(-0.5,9)(-0.5,12)}
\put(9,0){\put(-0.5,0){\line(0,1){12}}\put(0.5,0){\line(0,1){12}}}\put(9,0){\put(8.5,0){\line(0,1){12}}\put(9.5,0){\line(0,1){12}}}
}

\put(-100,132){\qbezier(0.5,0)(0.5,3)(3.5,4.5)\qbezier(6.5,7)(9.5,9)(9.5,12)\qbezier(-0.5,0)(-0.5,3)(2.5,5)\qbezier(5.5,7.5)(8.5,9)(8.5,12)
\qbezier(8.5,0)(8.5,3)(3.5,6)\qbezier(3.5,6)(-0.5,9)(-0.5,12)\qbezier(9.5,0)(9.5,3)(5.5,6)\qbezier(5.5,6)(0.5,9)(0.5,12)
\put(18,0){\put(-0.5,0){\line(0,1){12}}\put(0.5,0){\line(0,1){12}}}\put(-18,0){\put(8.5,0){\line(0,1){12}}\put(9.5,0){\line(0,1){12}}}
}

\put(-100,144){\qbezier(0.5,0)(0.5,3)(3.5,4.5)\qbezier(6.5,7)(9.5,9)(9.5,12)\qbezier(-0.5,0)(-0.5,3)(2.5,5)\qbezier(5.5,7.5)(8.5,9)(8.5,12)
\qbezier(8.5,0)(8.5,3)(3.5,6)\qbezier(3.5,6)(-0.5,9)(-0.5,12)\qbezier(9.5,0)(9.5,3)(5.5,6)\qbezier(5.5,6)(0.5,9)(0.5,12)
\put(18,0){\put(-0.5,0){\line(0,1){12}}\put(0.5,0){\line(0,1){12}}}\put(-18,0){\put(8.5,0){\line(0,1){12}}\put(9.5,0){\line(0,1){12}}}
}

\put(-100,156){
\put(-9,0){\qbezier(0.5,0)(0.5,3)(5.5,6)\qbezier(5.5,6)(9.5,9)(9.5,12)\qbezier(-0.5,0)(-0.5,3)(3.5,6)\qbezier(3.5,6)(8.5,9)(8.5,12)
\qbezier(9.5,0)(9.5,3)(6.5,5)\qbezier(3.5,7.5)(0.5,9)(0.5,12)\qbezier(8.5,0)(8.5,3)(5.5,4.5)\qbezier(2.5,7)(-0.5,9)(-0.5,12)}
\put(9,0){\put(-0.5,0){\line(0,1){12}}\put(0.5,0){\line(0,1){12}}}\put(9,0){\put(8.5,0){\line(0,1){12}}\put(9.5,0){\line(0,1){12}}}
}

\put(-10,230){\hbox{$k=26$}}
\put(-15,-20){\hbox{$P=0.99$}}
\put(-15,-35){\hbox{$\phi=0.29\pi$}}

\put(0,0){\qbezier(0.5,0)(0.5,3)(5.5,6)\qbezier(5.5,6)(9.5,9)(9.5,12)\qbezier(-0.5,0)(-0.5,3)(3.5,6)\qbezier(3.5,6)(8.5,9)(8.5,12)
\qbezier(9.5,0)(9.5,3)(6.5,5)\qbezier(3.5,7.5)(0.5,9)(0.5,12)\qbezier(8.5,0)(8.5,3)(5.5,4.5)\qbezier(2.5,7)(-0.5,9)(-0.5,12)
\put(18,0){\put(-0.5,0){\line(0,1){12}}\put(0.5,0){\line(0,1){12}}}\put(-18,0){\put(8.5,0){\line(0,1){12}}\put(9.5,0){\line(0,1){12}}}
}

\put(0,12){
\put(-9,0){\qbezier(0.5,0)(0.5,3)(3.5,4.5)\qbezier(6.5,7)(9.5,9)(9.5,12)\qbezier(-0.5,0)(-0.5,3)(2.5,5)\qbezier(5.5,7.5)(8.5,9)(8.5,12)
\qbezier(8.5,0)(8.5,3)(3.5,6)\qbezier(3.5,6)(-0.5,9)(-0.5,12)\qbezier(9.5,0)(9.5,3)(5.5,6)\qbezier(5.5,6)(0.5,9)(0.5,12)}
\put(9,0){\put(-0.5,0){\line(0,1){12}}\put(0.5,0){\line(0,1){12}}}\put(9,0){\put(8.5,0){\line(0,1){12}}\put(9.5,0){\line(0,1){12}}}
}
\put(0,24){
\put(-9,0){\qbezier(0.5,0)(0.5,3)(3.5,4.5)\qbezier(6.5,7)(9.5,9)(9.5,12)\qbezier(-0.5,0)(-0.5,3)(2.5,5)\qbezier(5.5,7.5)(8.5,9)(8.5,12)
\qbezier(8.5,0)(8.5,3)(3.5,6)\qbezier(3.5,6)(-0.5,9)(-0.5,12)\qbezier(9.5,0)(9.5,3)(5.5,6)\qbezier(5.5,6)(0.5,9)(0.5,12)}
\put(9,0){\put(-0.5,0){\line(0,1){12}}\put(0.5,0){\line(0,1){12}}}\put(9,0){\put(8.5,0){\line(0,1){12}}\put(9.5,0){\line(0,1){12}}}
}

\put(0,36){\qbezier(0.5,0)(0.5,3)(5.5,6)\qbezier(5.5,6)(9.5,9)(9.5,12)\qbezier(-0.5,0)(-0.5,3)(3.5,6)\qbezier(3.5,6)(8.5,9)(8.5,12)
\qbezier(9.5,0)(9.5,3)(6.5,5)\qbezier(3.5,7.5)(0.5,9)(0.5,12)\qbezier(8.5,0)(8.5,3)(5.5,4.5)\qbezier(2.5,7)(-0.5,9)(-0.5,12)
\put(18,0){\put(-0.5,0){\line(0,1){12}}\put(0.5,0){\line(0,1){12}}}\put(-18,0){\put(8.5,0){\line(0,1){12}}\put(9.5,0){\line(0,1){12}}}
}

\put(0,48){
\put(-9,0){\qbezier(0.5,0)(0.5,3)(3.5,4.5)\qbezier(6.5,7)(9.5,9)(9.5,12)\qbezier(-0.5,0)(-0.5,3)(2.5,5)\qbezier(5.5,7.5)(8.5,9)(8.5,12)
\qbezier(8.5,0)(8.5,3)(3.5,6)\qbezier(3.5,6)(-0.5,9)(-0.5,12)\qbezier(9.5,0)(9.5,3)(5.5,6)\qbezier(5.5,6)(0.5,9)(0.5,12)}
\put(9,0){\put(-0.5,0){\line(0,1){12}}\put(0.5,0){\line(0,1){12}}}\put(9,0){\put(8.5,0){\line(0,1){12}}\put(9.5,0){\line(0,1){12}}}
}
\put(0,60){
\put(-9,0){\qbezier(0.5,0)(0.5,3)(3.5,4.5)\qbezier(6.5,7)(9.5,9)(9.5,12)\qbezier(-0.5,0)(-0.5,3)(2.5,5)\qbezier(5.5,7.5)(8.5,9)(8.5,12)
\qbezier(8.5,0)(8.5,3)(3.5,6)\qbezier(3.5,6)(-0.5,9)(-0.5,12)\qbezier(9.5,0)(9.5,3)(5.5,6)\qbezier(5.5,6)(0.5,9)(0.5,12)}
\put(9,0){\put(-0.5,0){\line(0,1){12}}\put(0.5,0){\line(0,1){12}}}\put(9,0){\put(8.5,0){\line(0,1){12}}\put(9.5,0){\line(0,1){12}}}
}
\put(0,72){
\put(-9,0){\qbezier(0.5,0)(0.5,3)(3.5,4.5)\qbezier(6.5,7)(9.5,9)(9.5,12)\qbezier(-0.5,0)(-0.5,3)(2.5,5)\qbezier(5.5,7.5)(8.5,9)(8.5,12)
\qbezier(8.5,0)(8.5,3)(3.5,6)\qbezier(3.5,6)(-0.5,9)(-0.5,12)\qbezier(9.5,0)(9.5,3)(5.5,6)\qbezier(5.5,6)(0.5,9)(0.5,12)}
\put(9,0){\put(-0.5,0){\line(0,1){12}}\put(0.5,0){\line(0,1){12}}}\put(9,0){\put(8.5,0){\line(0,1){12}}\put(9.5,0){\line(0,1){12}}}
}
\put(0,84){
\put(-9,0){\qbezier(0.5,0)(0.5,3)(3.5,4.5)\qbezier(6.5,7)(9.5,9)(9.5,12)\qbezier(-0.5,0)(-0.5,3)(2.5,5)\qbezier(5.5,7.5)(8.5,9)(8.5,12)
\qbezier(8.5,0)(8.5,3)(3.5,6)\qbezier(3.5,6)(-0.5,9)(-0.5,12)\qbezier(9.5,0)(9.5,3)(5.5,6)\qbezier(5.5,6)(0.5,9)(0.5,12)}
\put(9,0){\put(-0.5,0){\line(0,1){12}}\put(0.5,0){\line(0,1){12}}}\put(9,0){\put(8.5,0){\line(0,1){12}}\put(9.5,0){\line(0,1){12}}}
}
\put(0,96){
\put(-9,0){\qbezier(0.5,0)(0.5,3)(3.5,4.5)\qbezier(6.5,7)(9.5,9)(9.5,12)\qbezier(-0.5,0)(-0.5,3)(2.5,5)\qbezier(5.5,7.5)(8.5,9)(8.5,12)
\qbezier(8.5,0)(8.5,3)(3.5,6)\qbezier(3.5,6)(-0.5,9)(-0.5,12)\qbezier(9.5,0)(9.5,3)(5.5,6)\qbezier(5.5,6)(0.5,9)(0.5,12)}
\put(9,0){\put(-0.5,0){\line(0,1){12}}\put(0.5,0){\line(0,1){12}}}\put(9,0){\put(8.5,0){\line(0,1){12}}\put(9.5,0){\line(0,1){12}}}
}

\put(0,108){\qbezier(0.5,0)(0.5,3)(3.5,4.5)\qbezier(6.5,7)(9.5,9)(9.5,12)\qbezier(-0.5,0)(-0.5,3)(2.5,5)\qbezier(5.5,7.5)(8.5,9)(8.5,12)
\qbezier(8.5,0)(8.5,3)(3.5,6)\qbezier(3.5,6)(-0.5,9)(-0.5,12)\qbezier(9.5,0)(9.5,3)(5.5,6)\qbezier(5.5,6)(0.5,9)(0.5,12)
\put(18,0){\put(-0.5,0){\line(0,1){12}}\put(0.5,0){\line(0,1){12}}}\put(-18,0){\put(8.5,0){\line(0,1){12}}\put(9.5,0){\line(0,1){12}}}
}
\put(0,120){\qbezier(0.5,0)(0.5,3)(3.5,4.5)\qbezier(6.5,7)(9.5,9)(9.5,12)\qbezier(-0.5,0)(-0.5,3)(2.5,5)\qbezier(5.5,7.5)(8.5,9)(8.5,12)
\qbezier(8.5,0)(8.5,3)(3.5,6)\qbezier(3.5,6)(-0.5,9)(-0.5,12)\qbezier(9.5,0)(9.5,3)(5.5,6)\qbezier(5.5,6)(0.5,9)(0.5,12)
\put(18,0){\put(-0.5,0){\line(0,1){12}}\put(0.5,0){\line(0,1){12}}}\put(-18,0){\put(8.5,0){\line(0,1){12}}\put(9.5,0){\line(0,1){12}}}
}
\put(0,132){\qbezier(0.5,0)(0.5,3)(3.5,4.5)\qbezier(6.5,7)(9.5,9)(9.5,12)\qbezier(-0.5,0)(-0.5,3)(2.5,5)\qbezier(5.5,7.5)(8.5,9)(8.5,12)
\qbezier(8.5,0)(8.5,3)(3.5,6)\qbezier(3.5,6)(-0.5,9)(-0.5,12)\qbezier(9.5,0)(9.5,3)(5.5,6)\qbezier(5.5,6)(0.5,9)(0.5,12)
\put(18,0){\put(-0.5,0){\line(0,1){12}}\put(0.5,0){\line(0,1){12}}}\put(-18,0){\put(8.5,0){\line(0,1){12}}\put(9.5,0){\line(0,1){12}}}
}
\put(0,144){\qbezier(0.5,0)(0.5,3)(3.5,4.5)\qbezier(6.5,7)(9.5,9)(9.5,12)\qbezier(-0.5,0)(-0.5,3)(2.5,5)\qbezier(5.5,7.5)(8.5,9)(8.5,12)
\qbezier(8.5,0)(8.5,3)(3.5,6)\qbezier(3.5,6)(-0.5,9)(-0.5,12)\qbezier(9.5,0)(9.5,3)(5.5,6)\qbezier(5.5,6)(0.5,9)(0.5,12)
\put(18,0){\put(-0.5,0){\line(0,1){12}}\put(0.5,0){\line(0,1){12}}}\put(-18,0){\put(8.5,0){\line(0,1){12}}\put(9.5,0){\line(0,1){12}}}
}

\put(90,230){\hbox{$k=26$}}
\put(85,-20){\hbox{$P=0.99$}}
\put(85,-35){\hbox{$\phi=-0.87\pi$}}

\put(100,0){\qbezier(0.5,0)(0.5,3)(5.5,6)\qbezier(5.5,6)(9.5,9)(9.5,12)\qbezier(-0.5,0)(-0.5,3)(3.5,6)\qbezier(3.5,6)(8.5,9)(8.5,12)
\qbezier(9.5,0)(9.5,3)(6.5,5)\qbezier(3.5,7.5)(0.5,9)(0.5,12)\qbezier(8.5,0)(8.5,3)(5.5,4.5)\qbezier(2.5,7)(-0.5,9)(-0.5,12)
\put(18,0){\put(-0.5,0){\line(0,1){12}}\put(0.5,0){\line(0,1){12}}}\put(-18,0){\put(8.5,0){\line(0,1){12}}\put(9.5,0){\line(0,1){12}}}
}

\put(100,12){
\put(-9,0){\qbezier(0.5,0)(0.5,3)(3.5,4.5)\qbezier(6.5,7)(9.5,9)(9.5,12)\qbezier(-0.5,0)(-0.5,3)(2.5,5)\qbezier(5.5,7.5)(8.5,9)(8.5,12)
\qbezier(8.5,0)(8.5,3)(3.5,6)\qbezier(3.5,6)(-0.5,9)(-0.5,12)\qbezier(9.5,0)(9.5,3)(5.5,6)\qbezier(5.5,6)(0.5,9)(0.5,12)}
\put(9,0){\put(-0.5,0){\line(0,1){12}}\put(0.5,0){\line(0,1){12}}}\put(9,0){\put(8.5,0){\line(0,1){12}}\put(9.5,0){\line(0,1){12}}}
}
\put(100,24){
\put(-9,0){\qbezier(0.5,0)(0.5,3)(3.5,4.5)\qbezier(6.5,7)(9.5,9)(9.5,12)\qbezier(-0.5,0)(-0.5,3)(2.5,5)\qbezier(5.5,7.5)(8.5,9)(8.5,12)
\qbezier(8.5,0)(8.5,3)(3.5,6)\qbezier(3.5,6)(-0.5,9)(-0.5,12)\qbezier(9.5,0)(9.5,3)(5.5,6)\qbezier(5.5,6)(0.5,9)(0.5,12)}
\put(9,0){\put(-0.5,0){\line(0,1){12}}\put(0.5,0){\line(0,1){12}}}\put(9,0){\put(8.5,0){\line(0,1){12}}\put(9.5,0){\line(0,1){12}}}
}
\put(100,36){
\put(-9,0){\qbezier(0.5,0)(0.5,3)(3.5,4.5)\qbezier(6.5,7)(9.5,9)(9.5,12)\qbezier(-0.5,0)(-0.5,3)(2.5,5)\qbezier(5.5,7.5)(8.5,9)(8.5,12)
\qbezier(8.5,0)(8.5,3)(3.5,6)\qbezier(3.5,6)(-0.5,9)(-0.5,12)\qbezier(9.5,0)(9.5,3)(5.5,6)\qbezier(5.5,6)(0.5,9)(0.5,12)}
\put(9,0){\put(-0.5,0){\line(0,1){12}}\put(0.5,0){\line(0,1){12}}}\put(9,0){\put(8.5,0){\line(0,1){12}}\put(9.5,0){\line(0,1){12}}}
}
\put(100,48){
\put(-9,0){\qbezier(0.5,0)(0.5,3)(3.5,4.5)\qbezier(6.5,7)(9.5,9)(9.5,12)\qbezier(-0.5,0)(-0.5,3)(2.5,5)\qbezier(5.5,7.5)(8.5,9)(8.5,12)
\qbezier(8.5,0)(8.5,3)(3.5,6)\qbezier(3.5,6)(-0.5,9)(-0.5,12)\qbezier(9.5,0)(9.5,3)(5.5,6)\qbezier(5.5,6)(0.5,9)(0.5,12)}
\put(9,0){\put(-0.5,0){\line(0,1){12}}\put(0.5,0){\line(0,1){12}}}\put(9,0){\put(8.5,0){\line(0,1){12}}\put(9.5,0){\line(0,1){12}}}
}

\put(100,60){\qbezier(0.5,0)(0.5,3)(3.5,4.5)\qbezier(6.5,7)(9.5,9)(9.5,12)\qbezier(-0.5,0)(-0.5,3)(2.5,5)\qbezier(5.5,7.5)(8.5,9)(8.5,12)
\qbezier(8.5,0)(8.5,3)(3.5,6)\qbezier(3.5,6)(-0.5,9)(-0.5,12)\qbezier(9.5,0)(9.5,3)(5.5,6)\qbezier(5.5,6)(0.5,9)(0.5,12)
\put(18,0){\put(-0.5,0){\line(0,1){12}}\put(0.5,0){\line(0,1){12}}}\put(-18,0){\put(8.5,0){\line(0,1){12}}\put(9.5,0){\line(0,1){12}}}
}
\put(100,72){\qbezier(0.5,0)(0.5,3)(3.5,4.5)\qbezier(6.5,7)(9.5,9)(9.5,12)\qbezier(-0.5,0)(-0.5,3)(2.5,5)\qbezier(5.5,7.5)(8.5,9)(8.5,12)
\qbezier(8.5,0)(8.5,3)(3.5,6)\qbezier(3.5,6)(-0.5,9)(-0.5,12)\qbezier(9.5,0)(9.5,3)(5.5,6)\qbezier(5.5,6)(0.5,9)(0.5,12)
\put(18,0){\put(-0.5,0){\line(0,1){12}}\put(0.5,0){\line(0,1){12}}}\put(-18,0){\put(8.5,0){\line(0,1){12}}\put(9.5,0){\line(0,1){12}}}
}
\put(100,84){\qbezier(0.5,0)(0.5,3)(3.5,4.5)\qbezier(6.5,7)(9.5,9)(9.5,12)\qbezier(-0.5,0)(-0.5,3)(2.5,5)\qbezier(5.5,7.5)(8.5,9)(8.5,12)
\qbezier(8.5,0)(8.5,3)(3.5,6)\qbezier(3.5,6)(-0.5,9)(-0.5,12)\qbezier(9.5,0)(9.5,3)(5.5,6)\qbezier(5.5,6)(0.5,9)(0.5,12)
\put(18,0){\put(-0.5,0){\line(0,1){12}}\put(0.5,0){\line(0,1){12}}}\put(-18,0){\put(8.5,0){\line(0,1){12}}\put(9.5,0){\line(0,1){12}}}
}
\put(100,96){\qbezier(0.5,0)(0.5,3)(3.5,4.5)\qbezier(6.5,7)(9.5,9)(9.5,12)\qbezier(-0.5,0)(-0.5,3)(2.5,5)\qbezier(5.5,7.5)(8.5,9)(8.5,12)
\qbezier(8.5,0)(8.5,3)(3.5,6)\qbezier(3.5,6)(-0.5,9)(-0.5,12)\qbezier(9.5,0)(9.5,3)(5.5,6)\qbezier(5.5,6)(0.5,9)(0.5,12)
\put(18,0){\put(-0.5,0){\line(0,1){12}}\put(0.5,0){\line(0,1){12}}}\put(-18,0){\put(8.5,0){\line(0,1){12}}\put(9.5,0){\line(0,1){12}}}
}

\put(100,108){
\put(-9,0){\qbezier(0.5,0)(0.5,3)(3.5,4.5)\qbezier(6.5,7)(9.5,9)(9.5,12)\qbezier(-0.5,0)(-0.5,3)(2.5,5)\qbezier(5.5,7.5)(8.5,9)(8.5,12)
\qbezier(8.5,0)(8.5,3)(3.5,6)\qbezier(3.5,6)(-0.5,9)(-0.5,12)\qbezier(9.5,0)(9.5,3)(5.5,6)\qbezier(5.5,6)(0.5,9)(0.5,12)}
\put(9,0){\put(-0.5,0){\line(0,1){12}}\put(0.5,0){\line(0,1){12}}}\put(9,0){\put(8.5,0){\line(0,1){12}}\put(9.5,0){\line(0,1){12}}}
}
\put(100,120){
\put(-9,0){\qbezier(0.5,0)(0.5,3)(3.5,4.5)\qbezier(6.5,7)(9.5,9)(9.5,12)\qbezier(-0.5,0)(-0.5,3)(2.5,5)\qbezier(5.5,7.5)(8.5,9)(8.5,12)
\qbezier(8.5,0)(8.5,3)(3.5,6)\qbezier(3.5,6)(-0.5,9)(-0.5,12)\qbezier(9.5,0)(9.5,3)(5.5,6)\qbezier(5.5,6)(0.5,9)(0.5,12)}
\put(9,0){\put(-0.5,0){\line(0,1){12}}\put(0.5,0){\line(0,1){12}}}\put(9,0){\put(8.5,0){\line(0,1){12}}\put(9.5,0){\line(0,1){12}}}
}
\put(100,132){
\put(-9,0){\qbezier(0.5,0)(0.5,3)(3.5,4.5)\qbezier(6.5,7)(9.5,9)(9.5,12)\qbezier(-0.5,0)(-0.5,3)(2.5,5)\qbezier(5.5,7.5)(8.5,9)(8.5,12)
\qbezier(8.5,0)(8.5,3)(3.5,6)\qbezier(3.5,6)(-0.5,9)(-0.5,12)\qbezier(9.5,0)(9.5,3)(5.5,6)\qbezier(5.5,6)(0.5,9)(0.5,12)}
\put(9,0){\put(-0.5,0){\line(0,1){12}}\put(0.5,0){\line(0,1){12}}}\put(9,0){\put(8.5,0){\line(0,1){12}}\put(9.5,0){\line(0,1){12}}}
}
\put(100,144){
\put(-9,0){\qbezier(0.5,0)(0.5,3)(3.5,4.5)\qbezier(6.5,7)(9.5,9)(9.5,12)\qbezier(-0.5,0)(-0.5,3)(2.5,5)\qbezier(5.5,7.5)(8.5,9)(8.5,12)
\qbezier(8.5,0)(8.5,3)(3.5,6)\qbezier(3.5,6)(-0.5,9)(-0.5,12)\qbezier(9.5,0)(9.5,3)(5.5,6)\qbezier(5.5,6)(0.5,9)(0.5,12)}
\put(9,0){\put(-0.5,0){\line(0,1){12}}\put(0.5,0){\line(0,1){12}}}\put(9,0){\put(8.5,0){\line(0,1){12}}\put(9.5,0){\line(0,1){12}}}
}

\put(100,156){\qbezier(0.5,0)(0.5,3)(3.5,4.5)\qbezier(6.5,7)(9.5,9)(9.5,12)\qbezier(-0.5,0)(-0.5,3)(2.5,5)\qbezier(5.5,7.5)(8.5,9)(8.5,12)
\qbezier(8.5,0)(8.5,3)(3.5,6)\qbezier(3.5,6)(-0.5,9)(-0.5,12)\qbezier(9.5,0)(9.5,3)(5.5,6)\qbezier(5.5,6)(0.5,9)(0.5,12)
\put(18,0){\put(-0.5,0){\line(0,1){12}}\put(0.5,0){\line(0,1){12}}}\put(-18,0){\put(8.5,0){\line(0,1){12}}\put(9.5,0){\line(0,1){12}}}
}
\put(100,168){\qbezier(0.5,0)(0.5,3)(3.5,4.5)\qbezier(6.5,7)(9.5,9)(9.5,12)\qbezier(-0.5,0)(-0.5,3)(2.5,5)\qbezier(5.5,7.5)(8.5,9)(8.5,12)
\qbezier(8.5,0)(8.5,3)(3.5,6)\qbezier(3.5,6)(-0.5,9)(-0.5,12)\qbezier(9.5,0)(9.5,3)(5.5,6)\qbezier(5.5,6)(0.5,9)(0.5,12)
\put(18,0){\put(-0.5,0){\line(0,1){12}}\put(0.5,0){\line(0,1){12}}}\put(-18,0){\put(8.5,0){\line(0,1){12}}\put(9.5,0){\line(0,1){12}}}
}
\put(100,180){\qbezier(0.5,0)(0.5,3)(3.5,4.5)\qbezier(6.5,7)(9.5,9)(9.5,12)\qbezier(-0.5,0)(-0.5,3)(2.5,5)\qbezier(5.5,7.5)(8.5,9)(8.5,12)
\qbezier(8.5,0)(8.5,3)(3.5,6)\qbezier(3.5,6)(-0.5,9)(-0.5,12)\qbezier(9.5,0)(9.5,3)(5.5,6)\qbezier(5.5,6)(0.5,9)(0.5,12)
\put(18,0){\put(-0.5,0){\line(0,1){12}}\put(0.5,0){\line(0,1){12}}}\put(-18,0){\put(8.5,0){\line(0,1){12}}\put(9.5,0){\line(0,1){12}}}
}
\put(100,192){\qbezier(0.5,0)(0.5,3)(3.5,4.5)\qbezier(6.5,7)(9.5,9)(9.5,12)\qbezier(-0.5,0)(-0.5,3)(2.5,5)\qbezier(5.5,7.5)(8.5,9)(8.5,12)
\qbezier(8.5,0)(8.5,3)(3.5,6)\qbezier(3.5,6)(-0.5,9)(-0.5,12)\qbezier(9.5,0)(9.5,3)(5.5,6)\qbezier(5.5,6)(0.5,9)(0.5,12)
\put(18,0){\put(-0.5,0){\line(0,1){12}}\put(0.5,0){\line(0,1){12}}}\put(-18,0){\put(8.5,0){\line(0,1){12}}\put(9.5,0){\line(0,1){12}}}
}
\put(100,204){\qbezier(0.5,0)(0.5,3)(3.5,4.5)\qbezier(6.5,7)(9.5,9)(9.5,12)\qbezier(-0.5,0)(-0.5,3)(2.5,5)\qbezier(5.5,7.5)(8.5,9)(8.5,12)
\qbezier(8.5,0)(8.5,3)(3.5,6)\qbezier(3.5,6)(-0.5,9)(-0.5,12)\qbezier(9.5,0)(9.5,3)(5.5,6)\qbezier(5.5,6)(0.5,9)(0.5,12)
\put(18,0){\put(-0.5,0){\line(0,1){12}}\put(0.5,0){\line(0,1){12}}}\put(-18,0){\put(8.5,0){\line(0,1){12}}\put(9.5,0){\line(0,1){12}}}
}

\put(190,230){\hbox{$k=33$}}
\put(185,-20){\hbox{$P=0.99$}}
\put(185,-35){\hbox{$\phi=-0.24\pi$}}
\put(200,0){\qbezier(0.5,0)(0.5,3)(5.5,6)\qbezier(5.5,6)(9.5,9)(9.5,12)\qbezier(-0.5,0)(-0.5,3)(3.5,6)\qbezier(3.5,6)(8.5,9)(8.5,12)
\qbezier(9.5,0)(9.5,3)(6.5,5)\qbezier(3.5,7.5)(0.5,9)(0.5,12)\qbezier(8.5,0)(8.5,3)(5.5,4.5)\qbezier(2.5,7)(-0.5,9)(-0.5,12)
\put(18,0){\put(-0.5,0){\line(0,1){12}}\put(0.5,0){\line(0,1){12}}}\put(-18,0){\put(8.5,0){\line(0,1){12}}\put(9.5,0){\line(0,1){12}}}
}
\put(200,12){\qbezier(0.5,0)(0.5,3)(5.5,6)\qbezier(5.5,6)(9.5,9)(9.5,12)\qbezier(-0.5,0)(-0.5,3)(3.5,6)\qbezier(3.5,6)(8.5,9)(8.5,12)
\qbezier(9.5,0)(9.5,3)(6.5,5)\qbezier(3.5,7.5)(0.5,9)(0.5,12)\qbezier(8.5,0)(8.5,3)(5.5,4.5)\qbezier(2.5,7)(-0.5,9)(-0.5,12)
\put(18,0){\put(-0.5,0){\line(0,1){12}}\put(0.5,0){\line(0,1){12}}}\put(-18,0){\put(8.5,0){\line(0,1){12}}\put(9.5,0){\line(0,1){12}}}
}
\put(200,24){\qbezier(0.5,0)(0.5,3)(5.5,6)\qbezier(5.5,6)(9.5,9)(9.5,12)\qbezier(-0.5,0)(-0.5,3)(3.5,6)\qbezier(3.5,6)(8.5,9)(8.5,12)
\qbezier(9.5,0)(9.5,3)(6.5,5)\qbezier(3.5,7.5)(0.5,9)(0.5,12)\qbezier(8.5,0)(8.5,3)(5.5,4.5)\qbezier(2.5,7)(-0.5,9)(-0.5,12)
\put(18,0){\put(-0.5,0){\line(0,1){12}}\put(0.5,0){\line(0,1){12}}}\put(-18,0){\put(8.5,0){\line(0,1){12}}\put(9.5,0){\line(0,1){12}}}
}

\put(200,36){
\put(-9,0){\qbezier(0.5,0)(0.5,3)(3.5,4.5)\qbezier(6.5,7)(9.5,9)(9.5,12)\qbezier(-0.5,0)(-0.5,3)(2.5,5)\qbezier(5.5,7.5)(8.5,9)(8.5,12)
\qbezier(8.5,0)(8.5,3)(3.5,6)\qbezier(3.5,6)(-0.5,9)(-0.5,12)\qbezier(9.5,0)(9.5,3)(5.5,6)\qbezier(5.5,6)(0.5,9)(0.5,12)}
\put(9,0){\put(-0.5,0){\line(0,1){12}}\put(0.5,0){\line(0,1){12}}}\put(9,0){\put(8.5,0){\line(0,1){12}}\put(9.5,0){\line(0,1){12}}}
}
\put(200,48){
\put(-9,0){\qbezier(0.5,0)(0.5,3)(3.5,4.5)\qbezier(6.5,7)(9.5,9)(9.5,12)\qbezier(-0.5,0)(-0.5,3)(2.5,5)\qbezier(5.5,7.5)(8.5,9)(8.5,12)
\qbezier(8.5,0)(8.5,3)(3.5,6)\qbezier(3.5,6)(-0.5,9)(-0.5,12)\qbezier(9.5,0)(9.5,3)(5.5,6)\qbezier(5.5,6)(0.5,9)(0.5,12)}
\put(9,0){\put(-0.5,0){\line(0,1){12}}\put(0.5,0){\line(0,1){12}}}\put(9,0){\put(8.5,0){\line(0,1){12}}\put(9.5,0){\line(0,1){12}}}
}
\put(200,60){
\put(-9,0){\qbezier(0.5,0)(0.5,3)(3.5,4.5)\qbezier(6.5,7)(9.5,9)(9.5,12)\qbezier(-0.5,0)(-0.5,3)(2.5,5)\qbezier(5.5,7.5)(8.5,9)(8.5,12)
\qbezier(8.5,0)(8.5,3)(3.5,6)\qbezier(3.5,6)(-0.5,9)(-0.5,12)\qbezier(9.5,0)(9.5,3)(5.5,6)\qbezier(5.5,6)(0.5,9)(0.5,12)}
\put(9,0){\put(-0.5,0){\line(0,1){12}}\put(0.5,0){\line(0,1){12}}}\put(9,0){\put(8.5,0){\line(0,1){12}}\put(9.5,0){\line(0,1){12}}}
}
\put(200,72){
\put(-9,0){\qbezier(0.5,0)(0.5,3)(3.5,4.5)\qbezier(6.5,7)(9.5,9)(9.5,12)\qbezier(-0.5,0)(-0.5,3)(2.5,5)\qbezier(5.5,7.5)(8.5,9)(8.5,12)
\qbezier(8.5,0)(8.5,3)(3.5,6)\qbezier(3.5,6)(-0.5,9)(-0.5,12)\qbezier(9.5,0)(9.5,3)(5.5,6)\qbezier(5.5,6)(0.5,9)(0.5,12)}
\put(9,0){\put(-0.5,0){\line(0,1){12}}\put(0.5,0){\line(0,1){12}}}\put(9,0){\put(8.5,0){\line(0,1){12}}\put(9.5,0){\line(0,1){12}}}
}

\put(200,84){\qbezier(0.5,0)(0.5,3)(3.5,4.5)\qbezier(6.5,7)(9.5,9)(9.5,12)\qbezier(-0.5,0)(-0.5,3)(2.5,5)\qbezier(5.5,7.5)(8.5,9)(8.5,12)
\qbezier(8.5,0)(8.5,3)(3.5,6)\qbezier(3.5,6)(-0.5,9)(-0.5,12)\qbezier(9.5,0)(9.5,3)(5.5,6)\qbezier(5.5,6)(0.5,9)(0.5,12)
\put(18,0){\put(-0.5,0){\line(0,1){12}}\put(0.5,0){\line(0,1){12}}}\put(-18,0){\put(8.5,0){\line(0,1){12}}\put(9.5,0){\line(0,1){12}}}
}
\put(200,96){\qbezier(0.5,0)(0.5,3)(3.5,4.5)\qbezier(6.5,7)(9.5,9)(9.5,12)\qbezier(-0.5,0)(-0.5,3)(2.5,5)\qbezier(5.5,7.5)(8.5,9)(8.5,12)
\qbezier(8.5,0)(8.5,3)(3.5,6)\qbezier(3.5,6)(-0.5,9)(-0.5,12)\qbezier(9.5,0)(9.5,3)(5.5,6)\qbezier(5.5,6)(0.5,9)(0.5,12)
\put(18,0){\put(-0.5,0){\line(0,1){12}}\put(0.5,0){\line(0,1){12}}}\put(-18,0){\put(8.5,0){\line(0,1){12}}\put(9.5,0){\line(0,1){12}}}
}
\put(200,108){\qbezier(0.5,0)(0.5,3)(3.5,4.5)\qbezier(6.5,7)(9.5,9)(9.5,12)\qbezier(-0.5,0)(-0.5,3)(2.5,5)\qbezier(5.5,7.5)(8.5,9)(8.5,12)
\qbezier(8.5,0)(8.5,3)(3.5,6)\qbezier(3.5,6)(-0.5,9)(-0.5,12)\qbezier(9.5,0)(9.5,3)(5.5,6)\qbezier(5.5,6)(0.5,9)(0.5,12)
\put(18,0){\put(-0.5,0){\line(0,1){12}}\put(0.5,0){\line(0,1){12}}}\put(-18,0){\put(8.5,0){\line(0,1){12}}\put(9.5,0){\line(0,1){12}}}
}
\put(200,120){\qbezier(0.5,0)(0.5,3)(3.5,4.5)\qbezier(6.5,7)(9.5,9)(9.5,12)\qbezier(-0.5,0)(-0.5,3)(2.5,5)\qbezier(5.5,7.5)(8.5,9)(8.5,12)
\qbezier(8.5,0)(8.5,3)(3.5,6)\qbezier(3.5,6)(-0.5,9)(-0.5,12)\qbezier(9.5,0)(9.5,3)(5.5,6)\qbezier(5.5,6)(0.5,9)(0.5,12)
\put(18,0){\put(-0.5,0){\line(0,1){12}}\put(0.5,0){\line(0,1){12}}}\put(-18,0){\put(8.5,0){\line(0,1){12}}\put(9.5,0){\line(0,1){12}}}
}

\put(200,132){
\put(-9,0){\qbezier(0.5,0)(0.5,3)(3.5,4.5)\qbezier(6.5,7)(9.5,9)(9.5,12)\qbezier(-0.5,0)(-0.5,3)(2.5,5)\qbezier(5.5,7.5)(8.5,9)(8.5,12)
\qbezier(8.5,0)(8.5,3)(3.5,6)\qbezier(3.5,6)(-0.5,9)(-0.5,12)\qbezier(9.5,0)(9.5,3)(5.5,6)\qbezier(5.5,6)(0.5,9)(0.5,12)}
\put(9,0){\put(-0.5,0){\line(0,1){12}}\put(0.5,0){\line(0,1){12}}}\put(9,0){\put(8.5,0){\line(0,1){12}}\put(9.5,0){\line(0,1){12}}}
}
\put(200,144){
\put(-9,0){\qbezier(0.5,0)(0.5,3)(3.5,4.5)\qbezier(6.5,7)(9.5,9)(9.5,12)\qbezier(-0.5,0)(-0.5,3)(2.5,5)\qbezier(5.5,7.5)(8.5,9)(8.5,12)
\qbezier(8.5,0)(8.5,3)(3.5,6)\qbezier(3.5,6)(-0.5,9)(-0.5,12)\qbezier(9.5,0)(9.5,3)(5.5,6)\qbezier(5.5,6)(0.5,9)(0.5,12)}
\put(9,0){\put(-0.5,0){\line(0,1){12}}\put(0.5,0){\line(0,1){12}}}\put(9,0){\put(8.5,0){\line(0,1){12}}\put(9.5,0){\line(0,1){12}}}
}
\put(200,156){
\put(-9,0){\qbezier(0.5,0)(0.5,3)(3.5,4.5)\qbezier(6.5,7)(9.5,9)(9.5,12)\qbezier(-0.5,0)(-0.5,3)(2.5,5)\qbezier(5.5,7.5)(8.5,9)(8.5,12)
\qbezier(8.5,0)(8.5,3)(3.5,6)\qbezier(3.5,6)(-0.5,9)(-0.5,12)\qbezier(9.5,0)(9.5,3)(5.5,6)\qbezier(5.5,6)(0.5,9)(0.5,12)}
\put(9,0){\put(-0.5,0){\line(0,1){12}}\put(0.5,0){\line(0,1){12}}}\put(9,0){\put(8.5,0){\line(0,1){12}}\put(9.5,0){\line(0,1){12}}}
}
\put(200,168){
\put(-9,0){\qbezier(0.5,0)(0.5,3)(3.5,4.5)\qbezier(6.5,7)(9.5,9)(9.5,12)\qbezier(-0.5,0)(-0.5,3)(2.5,5)\qbezier(5.5,7.5)(8.5,9)(8.5,12)
\qbezier(8.5,0)(8.5,3)(3.5,6)\qbezier(3.5,6)(-0.5,9)(-0.5,12)\qbezier(9.5,0)(9.5,3)(5.5,6)\qbezier(5.5,6)(0.5,9)(0.5,12)}
\put(9,0){\put(-0.5,0){\line(0,1){12}}\put(0.5,0){\line(0,1){12}}}\put(9,0){\put(8.5,0){\line(0,1){12}}\put(9.5,0){\line(0,1){12}}}
}

\put(200,180){\qbezier(0.5,0)(0.5,3)(5.5,6)\qbezier(5.5,6)(9.5,9)(9.5,12)\qbezier(-0.5,0)(-0.5,3)(3.5,6)\qbezier(3.5,6)(8.5,9)(8.5,12)
\qbezier(9.5,0)(9.5,3)(6.5,5)\qbezier(3.5,7.5)(0.5,9)(0.5,12)\qbezier(8.5,0)(8.5,3)(5.5,4.5)\qbezier(2.5,7)(-0.5,9)(-0.5,12)
\put(18,0){\put(-0.5,0){\line(0,1){12}}\put(0.5,0){\line(0,1){12}}}\put(-18,0){\put(8.5,0){\line(0,1){12}}\put(9.5,0){\line(0,1){12}}}
}
\put(200,192){\qbezier(0.5,0)(0.5,3)(5.5,6)\qbezier(5.5,6)(9.5,9)(9.5,12)\qbezier(-0.5,0)(-0.5,3)(3.5,6)\qbezier(3.5,6)(8.5,9)(8.5,12)
\qbezier(9.5,0)(9.5,3)(6.5,5)\qbezier(3.5,7.5)(0.5,9)(0.5,12)\qbezier(8.5,0)(8.5,3)(5.5,4.5)\qbezier(2.5,7)(-0.5,9)(-0.5,12)
\put(18,0){\put(-0.5,0){\line(0,1){12}}\put(0.5,0){\line(0,1){12}}}\put(-18,0){\put(8.5,0){\line(0,1){12}}\put(9.5,0){\line(0,1){12}}}
}
\put(200,204){\qbezier(0.5,0)(0.5,3)(5.5,6)\qbezier(5.5,6)(9.5,9)(9.5,12)\qbezier(-0.5,0)(-0.5,3)(3.5,6)\qbezier(3.5,6)(8.5,9)(8.5,12)
\qbezier(9.5,0)(9.5,3)(6.5,5)\qbezier(3.5,7.5)(0.5,9)(0.5,12)\qbezier(8.5,0)(8.5,3)(5.5,4.5)\qbezier(2.5,7)(-0.5,9)(-0.5,12)
\put(18,0){\put(-0.5,0){\line(0,1){12}}\put(0.5,0){\line(0,1){12}}}\put(-18,0){\put(8.5,0){\line(0,1){12}}\put(9.5,0){\line(0,1){12}}}
}




\end{picture}
\caption{Several examples of braidings that correspond to entangling operators with high fidelity.
\label{ent-braid}}
\end{figure}

These quantities of course depend on the parameters of the theory. Generally speaking, they depend on $A$ and $q$, or, in other words, on $N$ and $k$. Since in this section we discuss only $U_q(SU(2))$ case, only dependency on $k$ remains. For different values of $k$ different braids could give us high enough probability $\mathcal{P}$. We provide several examples of simple braids for arbitrarily chosen values of $k$ which possess high probability and non-trivial phase shift, see Fig.\ref{ent-braid}.


\section{Conclusion}

In this paper we suggested a realization of a two-qubit entangling operation in a topological quantum computer. For this we take pairs of anyons from one-qubit plat structures and entangle them as cables. This helps to greatly reduce the probability of system moving out of computational space, which is a scourge of operations in topological quantum computer. Moreover by carefully choosing the entangling of these cables for particular parameters of the theory, $k$ and $N$, the probability of system remaining in the computable space, which is fidelity of two-qubit operation in topological quantum computer, can be made sufficiently large. We show this by providing some examples in the $U_q(SU(2))$ case, see Fig.\ref{ent-braid}.

While throughout the paper we considered only two-qubit operations, their structure and the fact that they keep the computational space allows, obviously, to use such gates between any pair of qubits, thus providing one with multiqubit topological quantum computations.

The structure we discuss in this paper can also be generalized to the more general case of the $U_q(SU(N))$, however there are several complexifications on this way. Cables can carry not only trivial and adjoint representation, but also symmetric and anti-symmetric representation. Since in this case $[1]\neq \bar{[1]}$, the directions of the strands matter and depending on these directions in the cable, either $[1]\otimes [1]=[2]+[1,1]$, or $[1]\otimes \bar{[1]}=\emptyset+\text{adj}$ decomposition will occur. In the latter case, while the whole structure is the same as we discussed in section \ref{s:2braid}, the matrices will be of the size six-by-six \cite{UniRac}, rather than three-by-three in (\ref{R2r}) and (\ref{U2r}).

For the case of the parallel strands in the cable $[1]\otimes [1]=[2]+[1,1]$, the situation is more complex. Now there are no trivial cables, therefore cables always interact with each other. This means that for the same entangling of the cables there will be separate probabilities for each of the cases $I.$, $II.$, $III.$ and $IV.$ (if we define them in analogy to the (\ref{cases})), and the fidelity of the operation is defined by the worst of them. Also now there will be codependent phase shifts in the matrix, corresponding to the operation, therefore it should be checked that the resulting operation will indeed be entangling. Nevertheless there is no principal obstruction on the way of $U_q(SU(N))$ topological quantum computer. We leave this generalization for the future studies.

\section*{Acknowledgements}

We are grateful for very useful discussions with A.Anokhina, A.Belov, E.Lanina, A.Popolitov and A.Sleptsov.

This work was supported by the Russian Science Foundation grant No 23-71-10058.

\end{document}